\pdfoutput=1

\documentclass[12pt,a4paper]{article}

\usepackage{jheplike,ifpdf,tocloft,rotating,amsmath,mathrsfs,mathtools,xcolor}
\usepackage[all,cmtip]{xy}

\widowpenalty=500\clubpenalty=1000\unitlength=1mm\textheight=8.7in\hypersetup{pdftitle={},pdfcreator={},linkcolor=[rgb]{0.15,0.35,0.75},colorlinks=true,citecolor=[rgb]{0.675,0,0.2},urlcolor=[rgb]{0.15,0.35,0.65}}

\let\olditemize\itemize\renewcommand{\itemize}{\vspace{-2pt}\olditemize\setlength{\itemsep}{1pt}\setlength{\parskip}{0pt}\setlength{\parsep}{-0pt}}
\let\oldenumerate\enumerate\renewcommand{\enumerate}{\vspace{-4pt}\oldenumerate\setlength{\itemsep}{1pt}\setlength{\parskip}{0pt}\setlength{\parsep}{0pt}}

\newcommand{\eq}[1]{\vspace{-0.5pt}\begin{equation}#1\vspace{-0.5pt}\end{equation}}

\newcommand{\fwbox}[2]{\text{\makebox[#1][c]{$\hspace{-150pt}\displaystyle#2\hspace{-150pt}$}}}
\newcommand{\fwboxL}[2]{\text{\makebox[#1][l]{$#2$}}}
\newcommand{\fwboxR}[2]{\text{\makebox[#1][r]{$#2$}}}
\newcommand{\fig}[3]{\raisebox{#1}{\includegraphics[scale=#2]{#3}}}
\newcommand{\bigger}[1]{\raisebox{-0.95pt}{\scalebox{1.25}{$#1$}}}
\newcommand{\mi}{\raisebox{0.75pt}{\scalebox{0.75}{$\hspace{-2pt}\,-\,\hspace{-0.5pt}$}}}
\newcommand{\pl}{\raisebox{0.75pt}{\scalebox{0.75}{$\hspace{-2pt}\,+\,\hspace{-0.5pt}$}}}
\renewcommand{\bar}{\overline}

\renewcommand{\tilde}{\widetilde}
\renewcommand{\phi}{\varphi}
\newcommand{\ab}[1]{\langle #1\rangle}
\renewcommand{\sb}[1]{[ #1]}
\newcommand{\x}[2]{(#1,\hspace{-1pt}#2)}
\renewcommand{\u}[2]{(\hspace{-0.5pt}#1;\hspace{-1.5pt}#2\hspace{-0.5pt})}
\newcommand{\newcap}{\mathrm{\raisebox{0.75pt}{{$\,\bigcap\,$}}}}
\newcommand{\tcap}{\scalebox{1}{$\!\newcap\!$}}
\newcommand{\tncap}{\scalebox{0.8}{$\!\newcap\!$}}
\newcommand{\proj}[1]{\left[#1\right]}
\newcommand{\gramian}[2]{\Delta_{#1}^{\hspace{-2pt}#2}}
\definecolor{hblue}{rgb}{0,0,0.575}
\definecolor{hred}{rgb}{0.575,0.0,0.225}

\title{~\\[50pt]{\Huge \mbox{Rationalizing Loop Integration}}\\[-24pt]}
\author{\vspace{-10pt}Jacob~L.~Bourjaily}\emailAdd{bourjaily@nbi.ku.dk}
\author{\!\!,\,\,Andrew~J.~McLeod}\emailAdd{amcleod@nbi.ku.dk}
\author{\!\!,\,\,Matt~von~Hippel}\emailAdd{mvonhippel@nbi.ku.dk}
\author{\!\!,\,\,Matthias~Wilhelm}\emailAdd{matthias.wilhelm@nbi.ku.dk}
\affiliation{Niels Bohr International Academy and Discovery Center, Niels Bohr Institute,\\University of Copenhagen, Blegdamsvej 17, DK-2100, Copenhagen \O, Denmark}
\abstract{
We show that direct Feynman-parametric loop integration is possible for a large class of planar multi-loop integrals. Much of this follows from the existence of manifestly dual-conformal Feynman-parametric representations of planar loop integrals, and the fact that many of the algebraic roots associated with (e.g.\ Landau) leading singularities are automatically rationalized in momentum-twistor space---facilitating direct integration via partial fractioning. We describe how momentum twistors may be chosen non-redundantly to parameterize particular integrals, and how strategic choices of coordinates can be used to expose kinematic limits of interest. We illustrate the power of these ideas with many concrete cases studied through four loops and involving as many as eight particles. Detailed examples are included as ancillary files to this work's submission to the {\tt arXiv}. 
}
\preprint{}

\begin{document}
\maketitle\thispagestyle{empty}

\pagenumbering{roman}
\clearpage
\pagenumbering{arabic}
\vspace{-6pt}%
\section{Introduction and Overview}\label{sec:introduction}%
\vspace{-6pt}

In the study of scattering amplitudes in quantum field theory, several major research programs are built upon the premise that Feynman integrals are hard---hard enough to seek alternative approaches which bypass them altogether. Differential equation methods, which reframe Feynman integrals in terms of their kinematic derivatives~\mbox{\cite{Kotikov:1990kg,Remiddi:1997ny,MullerStach:2012mp,Henn:2013pwa,Lee:2014ioa,Meyer:2016slj}}, and bootstrap methods, which identify unique functions that match the expected properties of amplitudes or integrals~\cite{Dixon:2011pw,Dixon:2013eka,Dixon:2014voa,Caron-Huot:2016owq,Drummond:2014ffa,Dixon:2016nkn,Li:2016ctv,Almelid:2017qju,Chicherin:2017dob}, both spring from this philosophy---namely, that Feynman (parameter) integrals should be avoided at all costs. Similar considerations have motivated the development of advanced techniques for evaluating these integrals by transforming them into different integral representations~\cite{Gluza:2007rt,Blumlein:2014maa,Ochman:2015fho}.

What motivates this instinctual aversion? The na\"ive answer is that the problem of direct integration is open-ended---that, unlike differentiation, there is no algorithm for integrating arbitrarily complicated expressions. This is true, but somewhat facile: while generic Feynman integrals are expected to evaluate to periods of considerable complexity (and the next-to-simplest class of integrations are only starting to be understood~\cite{Adams:2017ejb,Remiddi:2017har,Broedel:2017kkb,Broedel:2017siw,Broedel:2018iwv,Bourjaily:2017bsb,Bourjaily:2018ycu}), there also exist infinite classes that are expected to evaluate to polylogarithms. In these cases, there {\it are} general algorithms available~\cite{Brown:2009ta,Panzer:2015ida}, which have even been implemented in convenient computer packages~\cite{Panzer:2014caa,Bogner:2014mha}.

Despite this, Feynman-parametric integration has not been pushed as far as other methods---at least not for multi-leg multi-loop processes. This is because integration into polylogarithms requires recasting denominators into a manifestly linear form in each variable, so that each integration can be converted into a polylogarithm term by term. This requires partial fractioning (potentially higher-order) polynomials in the denominator, which often results in algebraic quantities that themselves cannot be partial fractioned. In particular, once square (or higher) roots involving Feynman parameters appear, integration in these parameters can no longer be carried out using easily automated methods. 

These obstructions are well understood, insofar as they can be predicted and characterized via polynomial reduction algorithms~\cite{Brown:2008um,Brown:2009ta,Bogner:2013tia,Panzer:2015ida}. There is, however, no general method for resolving them, or even for determining when they can be avoided. In some instances, a judicious change of variables has proven sufficient to rationalize otherwise obstructive algebraic roots~\cite{Panzer:2014gra}; but in each of these cases, the solution seems highly tailored to the particular problem at hand. This suggests searching for more generally advantageous classes of coordinates for integration, as well as the development of techniques that allow coordinates to be tailored to specific problems.   

\newpage
\paragraph{Spiritus Movens}~\\[-14pt] 

A natural place to experiment with direct integration is planar maximally supersymmetric ($\mathcal{N}\!=\!4$) Yang-Mills theory (SYM), where many of the complications associated with generic quantum field theories disappear. In particular, only a relatively small number of integrals contribute to amplitudes---many of which do not require any regularization. Consider for example the well-known representation of the integrand for the two-loop $n$-point MHV amplitude \cite{Bourjaily:2015jna},
\eq{\fwboxR{0pt}{\mathcal{A}_n^{L=2,\text{MHV}}=\bigger{\displaystyle\sum_{a<b<c<d<a}}\hspace{-5pt}}\fig{-29.67pt}{1}{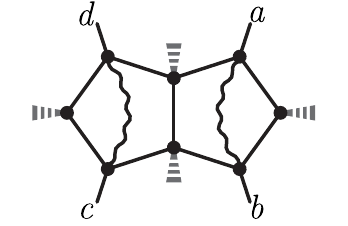},\!\label{double_pentagon_formula_for_mhv}}
wherein each term corresponds to a specific rational function in the external and loop momenta that takes the form
\eq{\begin{array}{@{}c@{}}\fwbox{80pt}{\fig{-29.67pt}{1}{mhv_general_double_pentagon}}\displaystyle\equiv\frac{\x{\ell_1}{N_1}\x{\ell_2}{N_2}}{\x{\ell_1}{a}\x{\ell_1}{a\pl\!1}\x{\ell_1}{b}\x{\ell_1}{b\pl1}\x{\ell_1}{\ell_2}\x{\ell_2}{c}\x{\ell_2}{c\pl1}\x{\ell_2}{d}\x{\ell_2}{d\pl1}}\,.\\[-12pt]\end{array}\label{mhv_double_pentagon_loop_space_integrand}\vspace{6pt}}
The factors $N_i$ are certain tensor numerators (indicated by the wavy lines in (\ref{double_pentagon_formula_for_mhv})) that are most easily defined in momentum-twistor space. But their precise form will not matter for the present discussion.

One important aspect of (\ref{double_pentagon_formula_for_mhv}) is that it cleanly separates the two-loop MHV amplitude integrand into (manifestly) infrared finite and divergent pieces \cite{Bourjaily:2015jna}. This allows us to discuss the `finite part' of the MHV two-loop amplitude, namely
\eq{\fwboxR{0pt}{\mathcal{A}_{n,\mathrm{fin}}^{L=2,\text{MHV}}=\bigger{\displaystyle\sum_{\substack{a+1<b<c\\c+1<d<a}}}\hspace{-5pt}}\fig{-29.67pt}{1}{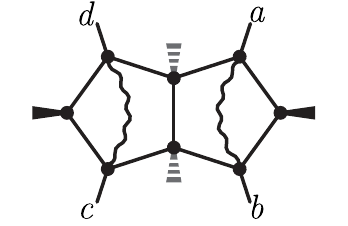}.\!\label{finite_part_of_mhv}}
The relationship between (\ref{finite_part_of_mhv}) and other characterizations of (the finite part of) the amplitude (e.g.\ the remainder function) is an interesting question, but not one we'll address here. Whatever the relation, it is clear that these integrals form an important part of finite `observables' related to MHV amplitudes at two loops. Moreover, as is obvious from the structure of (\ref{mhv_double_pentagon_loop_space_integrand}), an expression for the general case should capture {\it all} other cases via degenerations. 

A Feynman-parametric integral representation of (\ref{mhv_double_pentagon_loop_space_integrand}) that smoothly degenerates in these limits turns out to be reasonably straightforward to construct. Moreover, this representation can be made to depend explicitly on dual-conformal cross-ratios, via novel methods described in ref.\ \cite{conformalIntegration}. An integrand in this form is a natural candidate for direct integration. However, any attempt to carry out the integration over Feynman parameters is liable to encounter the obstruction highlighted above---namely, singularities can appear that are not rationally expressible in the cross-ratios (as is frequently the case with, e.g., Landau leading singularities).

As will be seen below, one source of such complicating roots is the existence of algebraic identities between multiplicatively independent cross-ratios. A good strategy for ameliorating this problem is to decompose all dual-conformally invariant cross-ratios into an independent set of momentum twistors, which rationalizes these roots. This strategy does more, though: by working in momentum-twistor space, it turns out that we automatically rationalize many of the otherwise algebraic roots that would have arisen in the course of direct integration (for instance at six points, where all multiplicatively independent cross-ratios are already algebraically independent).  (The value of rationalizing such kinematic roots was noted in \mbox{ref.\ \cite{Parker:2015cia}} as one of the key motivations for coordinate choices analogous to what we discuss here.)

The fact that (momentum-)twistor variables are useful for representing loop integrals such as (\ref{mhv_double_pentagon_loop_space_integrand}) has long been noted and exploited~\cite{Hodges:boxInt,Hodges:2009hk,ArkaniHamed:2010kv,Mason:2010pg,ArkaniHamed:2010gg,Parker:2015cia}. Indeed, virtually all known results regarding integrated multi-loop amplitudes in SYM make use of these variables. However, it is easy to see that simply going to momentum-twistor space introduces its own redundancies into the problem. Consider again the general double-pentagon integral (\ref{mhv_double_pentagon_loop_space_integrand}). A na\"ive representation in momentum-twistor space would use a pair of twistors for each massive corner, thus requiring twelve twistors for the general case. As reviewed below, the configuration space of twelve momentum twistors is 21$(\!=\!3\!\cdot\!12\mi15)$ dimensional. 

This is in fact considerably redundant. The number of degrees of freedom on which the double pentagon (\ref{mhv_double_pentagon_loop_space_integrand}) actually depends is not hard to count: it depends on exactly eight points in dual coordinates,
\eq{\fwbox{95pt}{\fig{-29.67pt}{1}{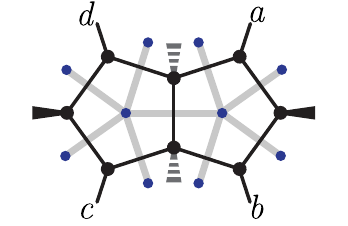}}\bigger{\Leftrightarrow}\fwbox{95pt}{\fig{-29.67pt}{1}{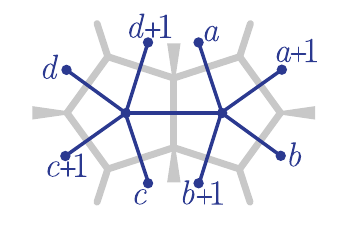}},}
four pairs of which are light-like separated. As reviewed in \mbox{section \ref{sec:kinematics}}, this means that the integral depends on only 13$(\!=\!4\!\cdot\!8\mi4\mi15)$ dual-conformal cross-ratios. It turns out that we can parametrize momentum twistors in terms of this reduced set of variables, by specializing to boundaries of the positive Grassmannian.

These, then, are our key ingredients: momentum twistors, parametrized in a non-redundant way. Thus equipped, we find that a broad class of kinematic square roots completely rationalize, enabling us to directly integrate several surprisingly complex classes of seven- and eight-point integrals through four loops.

Some of the integrals in \eqref{mhv_double_pentagon_loop_space_integrand} investigated here are being studied in parallel using the method of differential equations \cite{Henn:2018cdp}. These authors have determined the symbols of several of the two-loop integrals considered here, allowing for cross-checks of both results. In the future, we expect that our method for generating minimal parameterizations for such integrals will allow for further improvements to be made in the methods used there.

This paper is organized as follows. In \mbox{section \ref{sec:kinematics}}, we review how the kinematics of planar loop integrals can be encoded in terms of dual-momentum coordinates and dual-conformal cross-ratios. Using these, we describe a large class of Gramian determinants that are relevant to Feynman integrals. We then present the map between dual-momentum coordinates and momentum twistors, with an eye toward exploiting existing technology for concrete applications. These applications are laid out in \mbox{section \ref{sec:illustrations_of_integration}}, where we start with six-point integrals in \mbox{section \ref{subsec:hexagons}}, move on to seven-point integrals in \mbox{section \ref{subsec:heptagons}}, and finally discuss eight-point integrals in \mbox{section \ref{subsec:octagons_and_beyond}}---illustrating how appropriate kinematic parameterizations allow for the direct integration of many of the examples discussed. Along the way, we point out potential limitations of our methods---in particular, the appearance of algebraic roots that are not automatically rationalized by momentum twistors. We conclude in \mbox{section \ref{sec:conclusions}} with a discussion of directions for further research.

Finally, details of many of the concrete examples discussed in this work are included as ancillary files to this work's submission to the {\tt arXiv}. Due to file size restrictions, some of these expressions require {\sc Mathematica}'s {\tt Uncompress} function to unpack. The authors are happy to provide plain-text versions upon request.

\newpage\vspace{-0pt}%
\section[Rationalizing Variables for Planar Loop Integration]{Rationalizing Variables for Planar Loop Integration}\label{sec:kinematics}%
\vspace{-0pt}

\vspace{-2pt}
\subsection{Dual-Momentum Coordinates, Parameter Counts, and Gramians}\label{subsec:dual_momenta_and_conformal_invariance}%
\vspace{-0pt}

We are interested in planar Feynman integrals involving massless external particles. In order to trivialize momentum conservation, we introduce {\it dual-momentum} coordinates, associating the momentum $p_a$ of the $a$th external particle with the difference $p_a\!\equiv\!(x_{a+1}\mi x_a)$ (with cyclic labeling understood). Clearly, the map to dual coordinates is translationally invariant. In terms of these dual coordinates, Mandelstam invariants constructed out of consecutive sums of momenta may be expressed as
\eq{\x{a}{b}=\x{b}{a}\equiv(x_b\mi x_a)^2=(p_a\pl\!\ldots\pl p_{b-1})^2\,.\label{definition_of_two_brackets}}
This bracket is sometimes written `$x_{ab}^2$' in the literature. (It is worth mentioning that we often think of $\x{a}{b}$ as being defined in the embedding formalism---where it would be written `$X_a\!\cdot\!X_b$'. Our use of `$\x{a}{b}$' is designed to suggest such an inner product.) 

For a planar Feynman diagram, we may route the loop momenta according to the faces of the graph; assigning a dual point $x_{\ell_i}$ to each loop momentum (and exploiting translational invariance), each propagator involves either $\x{\ell_i}{a}\!\equiv\!(x_{\ell_i}\mi x_a)^2$ for some external point $x_a$, or $\x{\ell_i}{\ell_j}\!\equiv\!(x_{\ell_i}\mi x_{\ell_j})^2$ for those internal to the graph. It is easy to see that momentum conservation throughout the graph is automatically enforced. In this language, a loop integrand takes the form of a correlator associated with the (Poincar\'{e}-)dual of the Feynman graph. For example, the integral
\eq{I_{8,\text{A}}^{(2)}\equiv\fig{-29.67pt}{1}{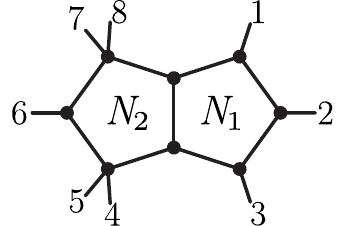}\bigger{\Leftrightarrow}\fig{-29.67pt}{1}{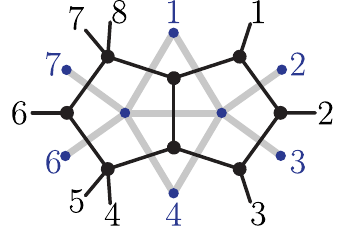}\bigger{\Leftrightarrow}\fig{-29.67pt}{1}{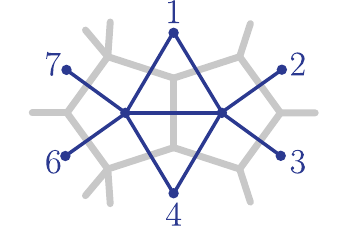}\label{octagon_a_dual_sequence_section_2}}
(discussed at length in \mbox{section \ref{subsec:octagons_and_beyond}}) would be expressed in dual-momentum space as
\eq{I_{8,\text{A}}^{(2)}\equiv\int\!\!\!d^4\ell_1d^4\ell_2\frac{\x{\ell_1}{N_1}\x{\ell_2}{N_2}}{\x{\ell_1}{1}\x{\ell_1}{2}\x{\ell_1}{3}\x{\ell_1}{4}\x{\ell_1}{\ell_2}\x{\ell_2}{4}\x{\ell_2}{6}\x{\ell_2}{7}\x{\ell_2}{1}}\,,\label{octagon_1_two_loop_definition}}
where definitions of the numerators $N_i$ are given in \mbox{appendix \ref{appendix:nonmhv_numerator_details}}.

It was noticed early on that (once canonically normalized) loop {\it integrands} such as (\ref{octagon_1_two_loop_definition}) were in fact conformally invariant in dual-momentum $x$-space \cite{Drummond:2006rz,Bern:2008ap,Drummond:2008aq}. Conformal invariance in dual-momentum coordinates is called {\it dual-conformal invariance}---sometimes `DCI' for short. It turns out that all infrared divergences (associated with integrals relevant to amplitudes in planar SYM) can be regulated without spoiling dual-conformal invariance \cite{Bourjaily:2013mma,conformalIntegration}, which proves that this symmetry survives as a symmetry for all infrared-finite quantities related to amplitudes in planar SYM. 

Although the Mandelstam invariants (\ref{definition_of_two_brackets}) are not DCI, cross-ratios constructed from them are:
\vspace{-0pt}\eq{\u{ab}{cd}\equiv\frac{\x{a}{b}\x{c}{d}}{\x{a}{c}\x{b}{d}}\,.\vspace{-0pt}\label{definition_of_cross_ratio_notation_in_x_space}}
Given a general configuration of $n$ points in $x$-space---that is, a configuration {\it not} involving light-like separated points---one can form $n(n\mi3)/2$ multiplicatively independent cross-ratios. If all pairs of neighboring points are light-like separated, this number becomes $n(n\mi5)/2$. However, in neither case does this count the number of algebraically independent cross-ratios, which is generally much smaller (and much easier to understand). 

The actual dimension of DCI kinematic invariants for a general configuration of $n$ dual points is given by $4\!\cdot\!n\mi15$: $4$ degrees of freedom per point $x_a$, minus the redundancy from the conformal group. If all pairs of neighboring points are light-like separated, the dimension is $3\!\cdot\!n\mi15$ (due to $n$ additional light-like constraints). In either case, the number of non-trivial relations satisfied by multiplicatively independent dual-conformal cross-ratios is
\vspace{-4pt}\eq{\text{\# of redundancies among DCI cross-ratios for $n$ particles: } (n\mi5)(n\mi6)/2\,,\vspace{-4pt}\label{overcounting_formula}}
from which we see that this redundancy first occurs for seven particles.

\vspace{-6pt}
\paragraph{Gramian Determinants and Algebraic Roots}\label{subsec:gramian_algebraic_relations}~\\[-14pt]%
\vspace{-0pt}

The easiest way to see that algebraic relations must be satisfied by the cross-ratios~\eqref{definition_of_cross_ratio_notation_in_x_space} is within the embedding formalism (see e.g.\ refs.\ \cite{Hodges:2010kq,SimmonsDuffin:2012uy,Bartels:2014mka}). In this context, it is obvious that $\x{a}{\cdot}$, viewed as an operator, spans a $6$-dimensional vector space. That is, the $n \times n$ Gramian matrix whose entries are built from the kinematic invariants $\x{a}{b}$, 
\vspace{-4pt}\eq{G\!\equiv\!\{G^a_{\phantom{a}b}\equiv\x{a}{b}\}\,,\vspace{-4pt}\label{definition_of_gramian_matrix_in_x_space}}
has rank at most six. In particular, all $7\!\times\!7$ minors of $G$ should vanish. Letting $G^A_{\phantom{A}B}$ denote the sub-matrix of $G$ involving rows $A$ and columns $B$,
\vspace{-4pt}\eq{\det\Big(G^{\{\fwbox{34pt}{a_1,\ldots,a_7}\}}_{\ \{\fwbox{34pt}{b_1,\ldots,b_7}\}}\Big)=0\,.\vspace{-4pt}\label{gramian_identities}}
These identities can always be normalized (by dividing by the leading term, say) so that they encode relations among (multiplicatively independent) dual-conformal cross-ratios. Unfortunately, even the simplest instance of a relation implied by (\ref{gramian_identities})---for seven particles---is too long to warrant writing here. We merely note that these relations among cross-ratios are at least quadratic, and their solutions involve algebraic roots depending on the set of cross-ratios chosen to be independent. 

Because such algebraic roots complicate much of the computational machinery involved in integration (or even analysis), it is worth enumerating at least one relevant class of these roots. They are associated with the $6\!\times\!6$ determinants of $G$,
\vspace{-4pt}\eq{\gramian{n}{A}\equiv\sqrt{\mi\det\!\big(G^{A}_{\phantom{A}A}\big)/\big(\x{a_1}{a_4}^2\x{a_2}{a_5}^2\x{a_3}{a_6}^2\big)}\quad\text{for } A\!=\!\{a_1,\ldots,a_6\}\,.\vspace{-4pt}\label{definition_of_gramian_6_determinants_in_x_space}}
For six particles, there is only one such Gramian root,
\eq{\gramian{6}{\{123456\}}=\sqrt{(1\mi u_1\mi u_2\mi u_3)^2\mi 4 u_1u_2u_3}\,,\label{six_particle_gram_det_section_2}}
where the dual-conformal cross-ratios $u_i$ are defined as
\eq{\fwbox{0pt}{\hspace{-30pt}u_1\!\equiv\!\u{13}{46}\!=\!\frac{\x{1}{3}\x{4}{6}}{\x{1}{4}\x{3}{6}},\, u_2\!\equiv\!\u{24}{51}\!=\!\frac{\x{2}{4}\x{1}{5}}{\x{2}{5}\x{1}{4}},\, u_3\!\equiv\!\u{35}{62}\!=\!\frac{\x{3}{5}\x{2}{6}}{\x{3}{6}\x{2}{5}}.}\label{uvw_defined_section_2}}
In general, there are $\binom{n}{6}$ such roots related to $6\!\times\!6$ minors of the Gramian matrix (\ref{definition_of_gramian_matrix_in_x_space}). Clearly, it would be advantageous to parameterize dual-conformal degrees of freedom in a way which rationalizes (at least) these algebraic roots. 

Working directly in the embedding formalism seems to achieve precisely this. Namely, we find that parametrizing external momenta in twistor space (which naturally realizes the embedding formalism) rationalizes all Gramian roots of the form (\ref{definition_of_gramian_6_determinants_in_x_space}). (However, as we shall see in section~\ref{subsubsec:octagonal_novelties_kinematic}, twistor space does not automatically rationalize all of the physically relevant algebraic roots encountered during loop integration.) 

Before reviewing how to parametrize our external kinematics in momentum-twistor space, we will take a slight detour to describe how some of these roots can arise in the process of loop integration via partial fractioning.

\vspace{-2pt}
\subsection{Algebraic Roots and Linear Reducibility}\label{subsec:partial_fractioning_roots}%
\vspace{-4pt}

Multiple polylogarithms (also known as `Goncharov' polylogarithms or hyperlogarithms \cite{goncharov:hyperlogs}) generalize logarithms to the space of iterated integrals taking the form
\eq{G_{a_1,\dots, a_n}(z) = \int_0^z \frac{dt}{t-a_1} G_{a_2,\dots, a_n}(z)\,, \quad G_{\fwboxL{27pt}{{\underbrace{0,\dots,0}_{p}}}}(z) = \frac{\log^p z}{p!} \,,\label{hyperlog_defn}}
where each variable $a_i$ can be an algebraic function of kinematic variables but not of integration variables~\mbox{\cite{Brown:2009qja,Goncharov:2009kx}}. Notably, while the denominators of these integrals are always linear in the integration variable, the integrals we are interested in generically involve denominators that are quadratic (or higher order) in Feynman parameters. These denominators may be partial fractioned, but at the possible cost of introducing algebraic roots. For instance, to integrate
\begin{align}
\int_0^\infty \frac{d \alpha}{\alpha^2\pl 2 f \alpha \pl g} &= \int_0^\infty \frac{d\alpha}{2 \sqrt{f^2\mi g}}\left( \frac{1}{\alpha \pl f \mi \sqrt{f^2\mi g}} \mi \frac{1}{\alpha \pl f \pl \sqrt{f^2\mi g}} \right) \nonumber \\
&= \frac{1}{2 \sqrt{f^2\mi g}} \log \left( \frac{f\mi\sqrt{f^2\mi g}}{f\pl\sqrt{f^2\mi g}} \right) , \label{eq: partial fractioning}
\end{align}
we are forced to introduce explicit factors of the roots of the polynomial \mbox{$\alpha^2\pl 2 f \alpha \pl g$}. Note that the integrals we consider in this paper are all normalized to have unit leading singularities. As such, they are expected to evaluate to `pure' polylogarithms. This requires extremely non-trivial cancellations between kinematic-dependent rational prefactors, which we see can in general be algebraic. This extraordinary property of these integrals can be highly non-obvious, indicating that we still have much to learn about how to organize polylogarithmic integration.

If all integrations can be carried out with the use of partial fractioning identities (and integration by parts to recognize total derivatives), a polylogarithmic integral is called {\it linearly reducible}~\cite{Brown:2008um,Brown:2009ta,Bogner:2013tia,Panzer:2015ida}. However, this does not always prove possible: in general, the functions $f$ and $g$ in (\ref{eq: partial fractioning}) may depend on other Feynman parameters, leading to integrands that depend on these parameters algebraically. This obstructs straightforward integration, since these Feynman parameters can no longer be partial fractioned so as to appear linearly in all denominators. Overcoming this obstacle generally requires finding a change of variables that rationalizes these algebraic roots (see e.g.\ ref.\ \cite{Panzer:2014gra}). (Note that such roots cannot be described by the Gramian, which is purely a function of kinematics.) We will encounter examples of this issue in sections~\ref{subsubsec:heptagon_c_section} and \ref{subsubsec:octagonal_novelties_algorithmic}, but we must leave the exploration of these issues to future work.

\vspace{-2pt}
\subsection{Momentum Twistors, Positivity, and Positroids}\label{subsec:momentum_twistors}%
\vspace{-4pt}

Although dual coordinates automatically enforce momentum conservation, they do not make manifest the masslessness of external particles. In dual coordinates, this corresponds to the non-trivial condition that \mbox{$p_a^2\!=\!\x{a}{a\pl1}\!=\!0$} for all $a$---that is, the constraint that neighboring dual coordinates are light-like separated. 

Andrew Hodges observed that the masslessness of external momenta would be easy to make manifest in the twistor space associated with dual-momentum $x$-coordinates \cite{Hodges:2009hk}. He called this {\it momentum-twistor} space. As with ordinary (spacetime) twistors, each point $x_a$ is associated with a line in momentum-twistor space; two points are null-separated in $x$-space iff their lines in twistor space intersect. A configuration of $n$ massless particles would then correspond to a collection of pairwise intersecting lines in momentum-twistor space---forming a polygon with $n$ vertices.

Thus, to describe $n$ massless particles in momentum-twistor space, we merely need $n$ arbitrarily distributed points in twistor space $z_a\!\in\!\mathbb{P}^3$ and to associate each dual coordinate $x_a$ with the `line' $(a)\!\equiv\!\mathrm{span}\{z_{a-1},z_a\}$. We often describe momentum twistors by four homogeneous coordinates---as $z_a\!\in\!\mathbb{C}^4/\text{GL}(1)$. In terms of these, `two lines intersect' iff the space spanned by the two lines is less than full rank (namely, 4). That is, lines $(a)$ and $(b)$ intersect iff $\det\{z_{a-1},z_a,z_{b-1},z_b\}\!=\!0$; from which it is trivial that $x_a\!\Leftrightarrow\!(a)$ and $x_{a+1}\!\Leftrightarrow\!(a\pl1)$ are light-like separated. We use angle-brackets to denote such determinants of (the homogeneous coordinates of) momentum twistors:
\eq{\ab{a\,b\,c\,d}\equiv\det\{z_a,z_b,z_c,z_d\}\,.\label{definition_of_four_bracket}}

It is not hard to see that conformal transformations in $x$-space translate to $\text{SL}(4)$ rotations in momentum-twistor space. From this, it is easy to understand the counting of independent dual-conformally invariant cross-ratios among $n$ massless particles: $(3n\mi15)$ is simply the dimension of the space of twistors $z_a\!\in\!\mathbb{P}^3$ modulo the action of $\text{SL}(4)$---which is $15$-dimensional. Thus, all four-brackets (\ref{definition_of_four_bracket}) are in fact dual-conformally invariant. 

In homogeneous coordinates, we may think of configurations of momentum twistors as being represented by $(4\!\times\!n)$ matrices $Z\!\equiv\!(z_1\cdots z_n)$ defined modulo the action of $\text{SL}(4)$ and a $\text{GL}(1)$ projective redundancy on each column of the matrix:
\eq{Z\equiv\big(z_1\,z_2\cdots z_n\big)/\text{SL}(4)\,,\quad\text{where } z_a\!\in\mathbb{P}^3(=\mathbb{C}^4/\text{GL}(1))\,.\label{momentum_twistor_matrix_notation}}
Equivalently, these are represented by points in the Grassmannian $G(4,n)$ of $4$-planes in $n$ dimensions, modulo the `torus action' of $\text{GL}(1)^{n-1}$. Thus, sets of momentum twistors represent points
\eq{Z\in G(4,n)/\mathrm{GL}(1)^{n-1}\,.\label{momentum_twistors_in_gr4n}}

Because (viewed homogeneously) twistor space is 4-dimensional, it is clear that the space of lines is $6(\!=\!\!\binom{4}{2}\!)$-dimensional. And indeed, any bi-twistor $\{z_a,z_b\}$ may be decomposed into a six-dimensional basis with rational coefficients. Thus, viewing $\x{a}{\cdot}$ as (something like) $\ab{(a\mi1 a)(\cdot)}$, it is clear that the Gramian matrix (\ref{definition_of_gramian_matrix_in_x_space}) will have rank $6$---trivializing all identities such as those in (\ref{gramian_identities}). 

Associating $\x{a}{b}$ with the four-bracket $\ab{a\mi1ab\mi1b}$ is a bit glib: for one thing, $\x{a}{b}$ is not dual-conformally invariant, whereas all four-brackets are (recall that the conformal group acts as $\text{SL}(4)$ in twistor space). Indeed, the explicit map between  $\x{a}{b}$ and $\ab{a\mi1a\,b\mi1b}$ requires reference to an explicitly-conformality-breaking `line at infinity' $(I_{\infty})$ 
\eq{\x{a}{b}=\frac{\ab{a\mi1a\,b\mi1b}}{\ab{a\mi1 a\,(I_{\infty})}\ab{b\mi1 b\,(I_{\infty})}}\,,\label{explicit_relation_between_2_and_4_brackets}}
where $(I_{\infty})$ is the line in momentum-twistor space corresponding to the point `$x_{\infty}$' in dual coordinates. Thus, the breaking of conformal invariance, which is fairly invisible in momentum-twistor space, is entirely associated with the correspondence (\ref{explicit_relation_between_2_and_4_brackets}) between twistors and $x$-space. However, while four-brackets are conformally invariant, they are not projectively invariant. (Recall that twistors should be viewed as points $z_a\!\in\!\mathbb{P}^3$.) This invariance is restored for {\it cross-ratios} in momentum-twistor space. In particular, the $x$-space cross-ratios defined in (\ref{definition_of_cross_ratio_notation_in_x_space}),
\eq{\u{ab}{cd}=\frac{\x{a}{b}\x{c}{d}}{\x{a}{c}\x{b}{d}}=\frac{\ab{a\mi1a\,b\mi1b}\ab{c\mi1c\,d\mi1d}}{\ab{a\mi1a\,c\mi1c}\ab{b\mi1b\,d\mi1d}}\,,\label{u_cross_ratios_in_twistor_space}}
are obviously both DCI and projectively invariant in momentum-twistor space. It is worth mentioning that the number of (multiplicatively independent) momentum-twistor cross-ratios can be (very) much larger than the numbers quoted for cross-ratios built exclusively from points in $x$-space. This does not really matter, however, as all redundancies among momentum-twistor four-brackets are captured by Pl\"{u}cker relations, which are always rational.

\vspace{-2pt}
\paragraph{The Euclidean Domain and (Twistor) Positivity}\label{subsec:euclidean_and_positive_regions}~\\[-14pt]%
\vspace{-4pt}

Loop integration requires a specification of the principal branch---the kinematic domain over which the integral is defined to be single-valued. From this domain of kinematics, other regions are accessible by analytic continuation in the usual way. 

For planar loop integrals, which depend exclusively on Mandelstam invariants constructed from consecutive sequences of momenta, \mbox{$\x{a}{b}\!=\!(p_a\pl\ldots\pl p_{b-1})^2$}, the obvious choice of principal branch corresponds to the condition that $\x{a}{b}\!\in\!\mathbb{R}_+$ for all $a,b$. This is often called the `Euclidean' domain. In the Euclidean domain, Feynman parametrization always results in a real-valued form whose only singularities reside at its boundaries, thereby manifesting the integral's single-valuedness.

Closely related to the Euclidean domain defined in $x$-space is the so-called `positive' domain of momentum twistors. A set of momentum twistors is said to be {\it positive} if $\ab{abcd}\!\in\!\mathbb{R}^+$ for all \mbox{$a\!<\!b\!<\!c\!<\!d$}. (We will often abuse terminology slightly and consider a configuration of twistors to be `positive' if merely $\ab{abcd}\!\geq\!0$, a condition more appropriately described as totally non-negative.) In homogeneous coordinates, this corresponds to the requirement that the $(4\!\times\!n)$ matrix $Z$ represents an element of the {\it positive Grassmannian} $G_+(4,n)$ \cite{P,lauren}. Configurations in the positive Grassmannian are known as {\it positroids}. 

It is easy to see that a positive configuration of momentum twistors can always define a point within the Euclidean domain in $x$-space. This is nearly trivial, but not entirely so. To see why, recall that the map between twistors and $x$-space requires reference to an infinity bi-twistor $(I_\infty)$; the signs of $\x{a}{b}$ are not determined entirely by the signs of $\ab{a\mi1a\,b\mi1 b}$ (which, by the way, are not always positive even within the positive domain, as $\x{1}{a}\!\sim\!\ab{n1\,a\mi1a}\!<\!0$ if $Z\!\in\!G_+(4,n)$). Nevertheless, it is not hard to show that for any positive configuration of momentum twistors, there always exists a bi-twistor $(I_\infty)$ for which $\ab{ab\,(I_\infty)}\!>\!0$ for all $a\!<\!b$. Such a choice of $(I_{\infty})$ ensures the positivity of all $\x{a}{b}$ via (\ref{explicit_relation_between_2_and_4_brackets}).  Thus, the positive domain is clearly a subspace of the Euclidean domain---and so should be equally well suited to defining the principal branch for loop integrals. 

Thus, the restriction to positive configurations of momentum twistors would seem at least well motivated through its connection to Euclidean kinematics. Although it remains unclear to which extent this is merely {\it technically} important, configurations of positive (or, more generally totally non-negative) twistors are extremely well understood mathematically (see e.g.\ \mbox{refs.\ \cite{P,lauren,knutson,ArkaniHamed:2009dg,ArkaniHamed:2012nw,Bourjaily:2012gy,positroidTwists}}). It is beyond the scope of this work to review this material here, but a few comments are in order to clarify the examples discussed in the next section. (A more thorough treatment of the positive-Grassmannian-motivated coordinate charts and symbol alphabets will be studied in a forthcoming work.)

\vspace{-2pt}
\paragraph{Canonical Coordinates on Positive Configurations of Twistors}\label{subsec:canonical_coordinates}~\\[-14pt]%
\vspace{-4pt}

Positive configurations of twistors, as examples of positroid varieties, are naturally endowed with a stratified set of boundaries on which different sets of four-brackets vanish. We expect these boundaries to play a privileged role in the representation of loop integrals (also in the non-planar case~\cite{Bourjaily:2016mnp}), but leave such exploration to future work. For now, these positroids merely furnish us with a convenient language in which to parametrize the (possibly lower-dimensional) configurations of momentum twistors that appear in the integrals under study. These positroid configurations come equipped with a wide variety of `canonical' coordinate charts, in which boundaries are linearly realized; cluster coordinates supply a familiar class of such charts. As virtually every example in this work will make use of canonical coordinates on positroids obtained via some planar bi-colored (`plabic') graph, it is worth describing the role played by the various {\it dramatis personae}. Developing the full theory behind these positroid structures would take us too far afield, so we keep this introduction brief and refer interested readers to \mbox{refs.\ \cite{P,ArkaniHamed:2012nw,Bourjaily:2012gy,positroidTwists}} for a more thorough discussion.

Positroid configurations are labeled by decorated permutations $\sigma\!:\!a\!\mapsto\!\sigma(a)$ among $n$ external twistors, which we label by the images $\{\sigma(1),\ldots,\sigma(n)\}$ with the convention that $\sigma(a)\!\geq\!a$ for all $a$. Geometrically, $\sigma(a)$ labels the {\it nearest} twistor `(cyclically) to the right of' $z_a$ for which
\eq{z_a\!\in\!\mathrm{span}\{z_{a+1},\ldots,z_{\sigma(a)}\}\,.}
For a generic configuration of twistors $Z\!\in\!G_+(4,n)$, $\sigma(a)\!=\!a\pl4$ for all $a$. More interestingly, the permutations describing the codimension-one boundaries of these configurations always correspond to transposing the images of two indices $\{a,b\}$, namely \mbox{$\sigma'\!=\!(ab)\!\circ\!\sigma$}, where $\sigma$ corresponds to the original configuration. However, not all such transpositions encode codimension-one boundaries; it can be shown that this is the case only when $a\!<\!b\leq \sigma(a)\!<\!\sigma(b)\!\leq\!a\pl n$, and when there exists no $a\!<\!c\!<\!b$ such that $\sigma(a)\!<\!\sigma(c)\!<\!\sigma(b)$. All boundaries of the positive region can be reached by some sequence of codimension-one boundaries. Interested readers should consult \mbox{e.g.\ refs.\ \cite{P,ArkaniHamed:2012nw,Bourjaily:2012gy}} for further details.

Canonical coordinate charts can be generated for the positroid configurations we will be interested in with the use of plabic graphs. Given a plabic graph, the positroid configuration it describes is determined by the permutation computed by the graph's left-right paths~\cite{ArkaniHamed:book}. A large number of plabic graphs are labeled by the same permutation, and any of these graphs can be used to generate canonical coordinates on the configuration of twistors labeled by this permutation (once the graph has been reduced by deleting internal bubbles). The coordinate charts generated by different graphs will in general be related by volume-preserving diffeomorphisms (which in special cases takes the form of cluster mutations).

Each individual plabic graph also comes equipped with multiple canonical coordinate charts, corresponding to different rules for associating coordinates to (oriented) edges and faces. Given a set of {\it edge} variables (generated by assigning a variable to each oriented edge in the graph), {\it face} variables can be defined as the product of all edge variables associated with the clockwise-aligned edges of each face, divided by all edge variables associated with the anticlockwise-aligned edges of the same face.\footnote{We emphasize that we refer to coordinates generated using this rule as {\it face} variables, although other coordinates can be assigned to the faces of a plabic graph. For instance, the cluster-$\mathcal{X}$ coordinates more typically discussed in the literature (see e.g.\ \mbox{refs.\ \cite{Golden:2013xva,Golden:2014xqa,Drummond:2014ffa,Dixon:2016nkn,DelDuca:2016lad,Drummond:2017ssj}}) are also associated to faces. These coordinates are non-trivially related to our face variables (which are also examples of cluster-$\mathcal{X}$ coordinates) by a `twist' operation \cite{positroidTwists}. See \mbox{appendix \ref{appendix:cluster_chart_comparisions}} for a concrete example of how these sets of variables are related.} Either edge or face variables can be used to parameterize a configuration according to the `boundary measurements' of a graph, as described in~\cite{ArkaniHamed:book}.

The only difference between the charts in which we are most interested and the direct output of, for example, the {\tt positroids} package \cite{Bourjaily:2012gy} is that we are primarily interested in charts on the {\it projective} Grassmannian, \mbox{$Z\!\in\!G_+(4,n)/\text{GL}(1)^{n-1}$}. The elimination of these $n\mi1$ relative redundancies can always be accomplished by setting $n\mi1$ of the face variables on the outer edge of the graph (the `Grassmannian necklace') to unity; the last face variable on this outer edge is then uniquely determined by the constraint that the product of all face variables be equal to unity (due to how these variables are constructed in terms of edge variables).

It may be helpful to illustrate boundary measurements in a concrete example. Consider the top-dimensional configuration of $n$ momentum twistors. The `lexicographically minimal' (lex-min) bridge-constructed plabic graph associated with the permutation $\sigma(a)\!=\!a\pl4$ (as generated by {\tt plabicGraph[Range[n]+4]} in the {\tt positroids} package~\cite{Bourjaily:2012gy}, for example) would be
\vspace{-4pt}\eq{\fig{-42.185pt}{1}{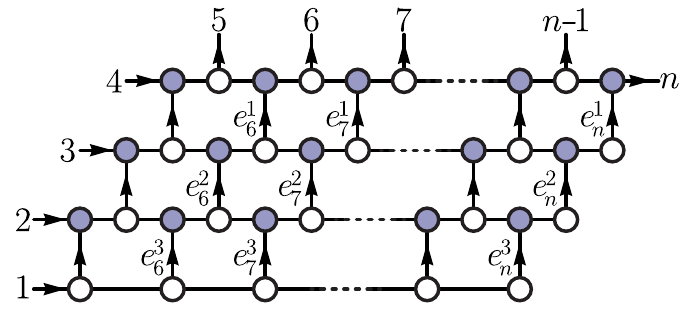}.\label{seed_plabic_graph}}
Here, we have set $(n\mi1)$ of the bridge variables associated with the outer faces to $1$. This chart corresponds to the boundary measurement matrix\footnote{Strictly speaking, the `boundary measurement matrix' for the graph drawn in (\ref{seed_plabic_graph}) would be the matrix obtained from (\ref{seed_edge_parameterization_for_top_cell_section2}) upon left-multiplication by $(z_1\,z_2\,z_3\,z_4)^{-1}$, which is just an $\text{SL}(4)$ transformation.}
\eq{Z^{(n)}_{\text{seed}}(\vec{e})\equiv\!\raisebox{-0pt}{$\left(\raisebox{54pt}{}\right.$}\begin{array}{@{$\;$}c@{$\;$}c@{$\;\;$}c@{$\;\;$}c@{$\;\;\;$}c@{$\;$}c@{$\;$}c@{$\;$}}z_1&z_2&z_3&z_4&\cdots&z_{a\in[4,n]}&\cdots\\[-1pt]\hline~\\[-14pt]
1&\displaystyle\hspace{-2pt}\sum_{4<k\leq n}\hspace{-6pt}e_k^3&\displaystyle\hspace{-2pt}\sum_{4<j<k\leq n}\hspace{-10pt}e_j^2e_k^3&\displaystyle\hspace{-2pt}\sum_{4<i<j<k\leq n}\hspace{-16pt}e_i^1e_j^2e_k^3&\cdots&\displaystyle\hspace{-2pt}\sum_{a<i<j<k\leq n}\hspace{-16pt}e_i^1e_j^2e_k^3&\cdots\\
0&1&\displaystyle\hspace{-2pt}\sum_{4<j\leq n}\hspace{-6pt}e_j^2&\displaystyle\hspace{-2pt}\sum_{4<i<j\leq n}\hspace{-10pt}e_i^1e_j^2&\cdots&\displaystyle\hspace{-2pt}\sum_{a<i<j\leq n}\hspace{-10pt}e_i^1e_j^2&\cdots\\
0&0&1&\displaystyle\hspace{-2pt}\sum_{4<i\leq n}\hspace{-6pt}e_i^1&\cdots&\displaystyle\hspace{-2pt}\sum_{a<i\leq n}\hspace{-6pt}e_i^1&\cdots\\
0&0&0&1&\cdots&1&\cdots\\~\end{array}\raisebox{-0pt}{$\left.\raisebox{54pt}{}\right)$}\,,\label{seed_edge_parameterization_for_top_cell_section2}\vspace{-20pt}}
where $e_5^i\!\equiv\!1$. It is easy to confirm that $e_a^i\!\in\!\mathbb{R}_+$ ensures that $Z^{(n)}_{\text{seed}}$ is positive. We use this chart in several examples in \mbox{section \ref{sec:illustrations_of_integration}}, if only as a convenient reference chart. 

However, as mentioned already in the introduction, we are often interested in loop integrals that depend on fewer dual coordinates than the number of external legs on the graph. In such cases, the number of $3n\mi15$ degrees of freedom associated with \mbox{$G_+(4,n)/\text{GL}(1)^{n-1}$} is much larger than the number we actually need. Correspondingly, we now turn to the language of the positroid stratification (permutations, plabic graphs, etc.), as it provides a natural way to parameterize coordinates in momentum-twistor space specifically tailored to a given integral. 

\vspace{-2pt}
\paragraph{Eliminating Redundancies in Momentum-Twistor Space}\label{subsec:eliminating_twistor_redundancies}~\\[-14pt]%
\vspace{-4pt}

Perhaps the easiest way to understand the redundancies involved in configurations of $n$ momentum twistors is to consider the following sequence of two-loop integral topologies: 
\eq{\hspace{-14pt}\fig{-29.67pt}{1}{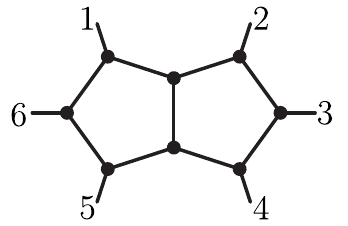}\hspace{-0pt}\bigger{\Rightarrow}\hspace{-0pt}\fig{-29.67pt}{1}{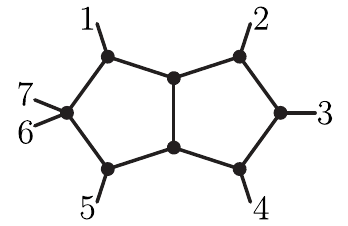}\hspace{-0pt}\bigger{\Rightarrow}\hspace{-0pt}\fig{-29.67pt}{1}{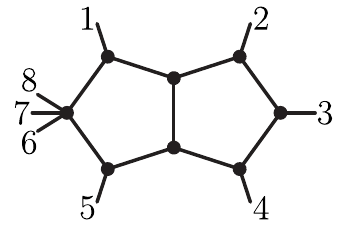}\hspace{-0pt}\bigger{\Rightarrow}\ldots\,.\label{adding_masses_to_legs_sequence}}
Although the number of external legs grows arbitrarily, the set of dual coordinates on which these integrals would depend barely changes. Indeed, regardless of the multiplicity, these integrals' denominators depend exclusively on six dual points. For example,
\eq{\hspace{-0pt}\fig{-29.67pt}{1}{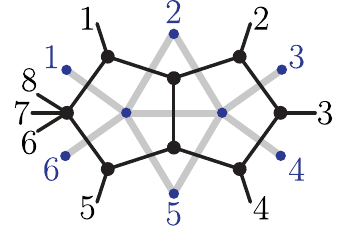}\bigger{\Leftrightarrow}\fig{-29.67pt}{1}{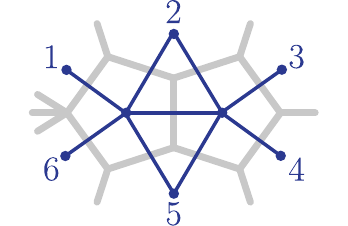}.\label{dual_equivalence_figure}}
Indeed, the only substantive difference among the integrals in (\ref{adding_masses_to_legs_sequence}) is that the first depends on six points $\{x_1,\ldots,x_6\}$ with all pairs of neighboring points light-like separated---{\it including} the pair $\{x_6,x_1\}$; while for all the others, the dual coordinates $x_1$ and $x_6$ should be understood to be in a general configuration relative to each other. As discussed previously, this means that the first integral in (\ref{adding_masses_to_legs_sequence}) should depend on 3 (dual-conformal) degrees of freedom, while all the others in the infinite sequence would depend on 4. The fact that the third (and further) integrals in the sequence (\ref{adding_masses_to_legs_sequence}) all amount to re-labeling the second case is semi-obvious. 

What is more interesting is the difference between the first two integrals in (\ref{adding_masses_to_legs_sequence}):
\eq{\hspace{-14pt}\fig{-29.67pt}{1}{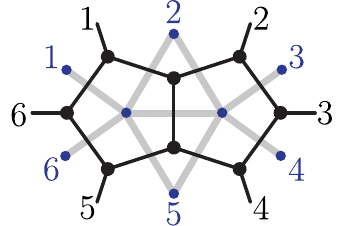}\hspace{-0pt}\;\text{vs.}\;\hspace{-0pt}\fig{-29.67pt}{1}{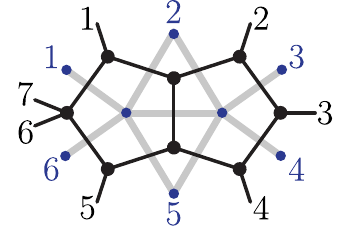}\,.\label{six_to_seven_massive_sequence}}
Na\"ively, momentum-twistor space would parameterize the second integral above in terms of $Z\!\in\!G_+(4,7)$---a space of dimension 6 (after projectivization). However, it is obvious that the second integral is independent of the dual point $x_7$, and that there should only be 4 degrees of freedom. 

(This is basically identical to what happens when we represent a massive particle's momentum in terms of a pair of massless momenta: the four degrees of freedom required for the massive particle would be represented by $2\!\times\!3$ degrees of freedom; which over-counts the right number by 2---the same degree of redundancy as seen in momentum-twistor space.)

Let us now describe how a configuration in twistor space that is independent of the point $x_7$ can be defined. Being independent of $x_7$ implies that a quantity does not depend on the line $(67)$, while it may still depend on the lines \mbox{$(71),(12),\ldots(56)$}, which encode the dual points \mbox{$x_1,x_2,\ldots,x_6$}. Any direct dependence on the line $(67)$ would be avoided if $(67)$ were in fact required to be the line `$(456)\tncap(712)$'---that is, the line spanned by the intersections of the planes $(456)$ and $(712)$, which can be can be represented more explicitly as
\eq{\begin{split}
({\color{hred}abc})\tcap({\color{hblue}def})&\equiv\mathrm{span}\{z_{\color{hred}a},z_{\color{hred}b},z_{\color{hred}c}\}\tcap\,\mathrm{span}\{z_{\color{hblue}d},z_{\color{hblue}e},z_{\color{hblue}f}\}\\
&=({\color{hred}ab})\ab{{\color{hred}c}\,{\color{hblue}def}}\pl({\color{hred}bc})\ab{{\color{hred}a}\,{\color{hblue}def}}\pl({\color{hred}ca})\ab{{\color{hred}b}\,{\color{hblue}def}}\\
&=\ab{{\color{hred}abc}\,{\color{hblue}d}}({\color{hblue}ef})\pl\ab{{\color{hred}abc}\,{\color{hblue}e}}({\color{hblue}fd})\pl\ab{{\color{hred}abc}\,{\color{hblue}f}}({\color{hblue}de})\,.\end{split}\label{definition_of_intersection_of_planes}}
From the definition above, it is easy to see that replacing $(67)$ with $(456)\tncap(712)$ would express it entirely in terms of the lines corresponding to the dual points $x_1,\ldots,x_6$. In terms of active transformations, this would be achieved by shifting $z_7\!\mapsto\!(71)\tncap(456)$ and $z_6\!\mapsto\!(56)\tncap(712)$. It is easy to see that these transformations leave the relevant lines $(56),(71),\ldots$ unchanged, while eliminating any dependence on the line $(67)$. 

If the geometric story is not sufficiently intuitive, it is worth mentioning that the constraints above are equivalent to the requirements that $\ab{{\color{hred}67}\,12}\!=\!\ab{45\,{\color{hred}67}}\!=\!0$. In $x$-space, these two conditions simply translate to the constraints that \mbox{$\x{{\color{hred}7}}{2}\!=\!\x{5}{{\color{hred}7}}\!=\!0$}, which is clearly something that can be imposed without any loss of generality for any integral that does not depend on $x_7$.

In terms of the positive Grassmannian, the generic configuration of momentum twistors would be labeled by the permutation \mbox{$\sigma\!=\!\{5,6,7,8,9,10,11\}$}---which is to say that $\sigma$ is the permutation \mbox{$\sigma\!:\!a\!\mapsto\!a\pl4$}. Setting $\ab{45\,{\color{hred}67}}\!=\!0$ would result in a configuration labeled by $\sigma'\!\equiv\![34]\!\circ\!\sigma\!=\!\{5,6,{\color{hred}8},{\color{hred}7},9,10,11\}$; upon additionally setting $\ab{{\color{hred}67}\,12}\!=\!0$, the configuration would be labeled by $\sigma''\!\equiv\![56]\!\circ\!\sigma'\!=\!\{5,6,8,7,{\color{hred}10},{\color{hred}9},11\}$. Thus, the codimension-two configuration labeled by $\sigma''$ would provide a four-dimensional parameterization of momentum twistors tailored to an integral such as the second integral drawn in (\ref{six_to_seven_massive_sequence}). 

Although fairly trivial, it is worth mentioning that the difference between
\eq{\hspace{-0pt}\fig{-29.67pt}{1}{hexagon_mass_sequence_1_bare}\hspace{-0pt}\;\text{and}\;\hspace{-0pt}\fig{-29.67pt}{1}{hexagon_mass_sequence_2_bare}\label{adding_masses_to_legs_sequence_2}}
can be understood as going from a (restricted) configuration in $G_+(4,7)$ to one in $G_+(4,8)$---which in this case does not introduce any new degrees of freedom. It should be obvious that the second integral in (\ref{adding_masses_to_legs_sequence_2}) is independent of twistor $z_7$. There are in fact two possible ways to realize such independence: either $z_7$ can be made proportional to $z_6$ or to $z_8$. In either case, the two points correspond to the same point in twistor space ($\mathbb{P}^3$). Geometrically, this would correspond to a plabic graph in which a pair of external legs (in this case, either the pair $\{6,7\}$ or $\{7,8\}$) were both connected to a white vertex. This has the interpretation (via boundary measurements) that all the `columns' attached to the vertex are proportional to one another.

Let us conclude this discussion with one final example (of particular relevance to MHV amplitudes). Recall from the introduction that the most general (finite) integral relevant to MHV amplitudes at two loops involves at least twelve external particles:
\vspace{4pt}\eq{\fwbox{0pt}{\fig{-29.67pt}{1}{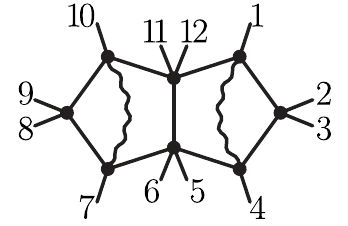}\bigger{\Leftrightarrow}\fig{-29.67pt}{1}{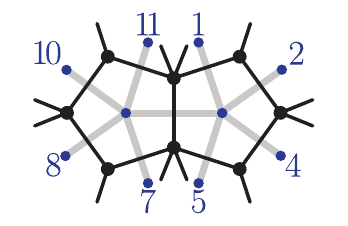}\bigger{\Leftrightarrow}\fig{-29.67pt}{1}{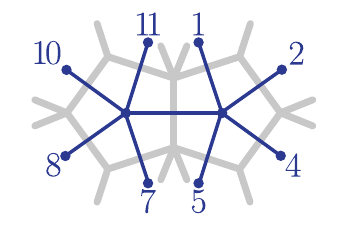}\fwboxL{0pt}{.}}\label{twelve_point_mhv_dual_sequence}}
As already mentioned, a general configuration of twelve momentum twistors is 21-dimensional. However, it is obvious that the integral (\ref{twelve_point_mhv_dual_sequence}) depends on only eight dual coordinates of which (at least)  four pairs are light-like separated. It is not hard to see that such a configuration (in dual-momentum space) should be merely 13-dimensional. 

Following the same strategy described above to make manifest the fact that the integral (\ref{twelve_point_mhv_dual_sequence}) is independent of the dual points \mbox{$\{x_3,x_6,x_9,x_{12}\}$}, we may impose the following eight conditions: 
\eq{\x{1}{{\color{hred}3}}\!=\!\x{{\color{hred}3}}{5}\!=\!\x{4}{{\color{hred}6}}\!=\!\x{{\color{hred}6}}{8}\!=\!\x{7}{{\color{hred}9}}\!=\!\x{{\color{hred}9}}{11}\!=\!\x{10}{{\color{hred}12}}\!=\!\x{{\color{hred}12}}{2}\!=\!0.\label{eight_constraints_in_x_space}}
In momentum-twistor space, the constraints (\ref{eight_constraints_in_x_space}) correspond to 
\eq{\fwbox{0pt}{\hspace{-30pt}\ab{1\hspace{-0.75pt}2\hspace{1pt}1\,{\color{hred}23}}\!\!=\!\!\ab{{\color{hred}23}\,45}\!\!=\!\!\ab{34\,{\color{hred}56}}\!\!=\!\!\ab{{\color{hred}56}\,78}\!\!=\!\!\ab{67\,{\color{hred}89}}\!\!=\!\!\ab{{\color{hred}89}\,1\hspace{-0.75pt}0\hspace{1pt}1\hspace{-0.75pt}1}\!\!=\!\!\ab{9\hspace{1pt}1\hspace{-0.75pt}0\,{\color{hred}1\hspace{-0.75pt}1\hspace{1pt}1\hspace{-0.75pt}2}}\!\!=\!\!\ab{{\color{hred}1\hspace{-0.75pt}1\hspace{1pt}1\hspace{-0.75pt}2}\,12}\!\!=\!\!0.}\label{eight_constraints_in_g412}}
It turns out that there are 16 codimension-8 configurations in $G_+(4,12)$ which satisfy the constraints (\ref{eight_constraints_in_g412}). (These are easily found using the {\tt positroids} package \cite{Bourjaily:2012gy}.) One of these is labeled by the permutation $\sigma\!=\!\{7,5,6,10,8,9,13,11,12,16,14,15\}$, 
a parameterization of which would be obtainable as boundary measurements of the plabic graph
\eq{\fwboxR{0pt}{Z\;\;\bigger{\Leftrightarrow}\;}\fig{-42.185pt}{1}{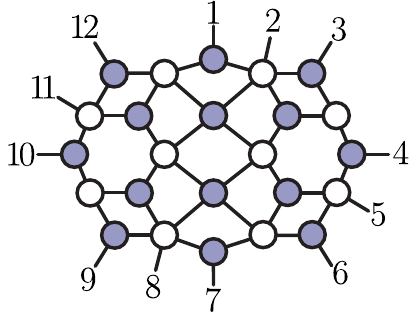}\fwboxL{0pt}{\,.}\label{plabic_graph_for_twelve_point_mhv_double_pentagon}}

This can be generalized to all multiplicity for the integrals of interest, using 
\eq{\fwbox{95pt}{\fig{-29.67pt}{1}{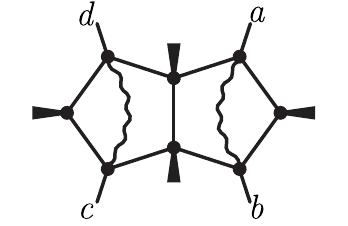}}\bigger{\Leftrightarrow}\fig{-42.185pt}{1}{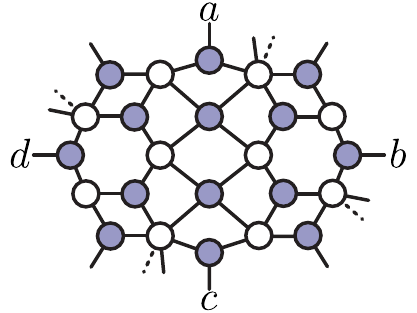},\label{plabic_graph_for_mhv_double_pentagon_chart}}
which would parameterize a configuration of momentum twistors labeled by
\eq{\fwbox{0pt}{\hspace{-26pt}\sigma\!:\!\!\Bigg(\!\begin{array}{@{}c@{$\;$}c@{$\;$}c@{$\;$}c@{$\;$}c@{$\;$}c@{$\;$}c@{$\;$}c@{$\;$}c@{$\;$}c@{$\;$}c@{$\;$}c@{$\;$}c@{$\;$}c@{$\;$}c@{$\;$}c@{$\;$}c@{$\;$}c@{$\;$}c@{$\;$}c@{$\;$}c@{$\;$}c@{$\;$}c@{$\;$}c@{}}a\mi2&a\mi\!1&a&\cdots&b\mi2&b\mi\!1&b&\cdots&c\mi2&c\mi\!1&c&\cdots&d\mi2&d\mi\!1&d&\cdots\\[-6pt]
\downarrow&\downarrow&\downarrow&{}&\downarrow&\downarrow&\downarrow&{}&\downarrow&\downarrow&\downarrow&&\downarrow&\downarrow&\downarrow\\[-4pt]
a\pl\!1&b\mi\!1&c&\cdots&b\pl\!1&c\mi\!1&d&\cdots&c\pl\!1&d\mi\!1&a&\cdots&d\pl\!1&a\mi\!1&b&\cdots\end{array}\!\!\Bigg)\!\text{, else }\sigma\!:\!\!\Bigg(\!\begin{array}{@{}c@{}}e\\[-6pt]
\downarrow\\[-4pt]
e\pl\!1\end{array}\!\Bigg)\!.}\label{general_mhv_cell_permutation}}
%

\newpage\vspace{-6pt}
\section[Illustrations of Loop Integration in Twistor Space]{Illustrations of Loop Integration in Twistor Space}\label{sec:illustrations_of_integration}%
\vspace{-6pt}

\vspace{-2pt}
\subsection{Hexagon Integrals---Warmup and Review/Overview}\label{subsec:hexagons}%
\vspace{-2pt}

There is only a single two-loop integral topology for six particles that can be rendered infrared finite by a suitable choice of tensor numerators, namely
\eq{\hspace{-24pt}\fwbox{90pt}{\fig{-29.67pt}{1}{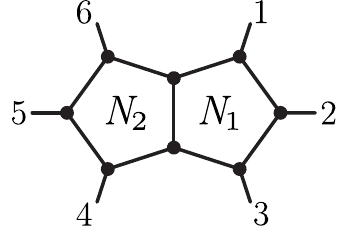}} \newline \equiv\!\!\int\!\!\!d^{2\times4}\vec{\ell}\frac{\x{\ell_1}{N_1}\x{1}{4}\x{\ell_2}{N_2}}{\x{\ell_1}{1}\x{\ell_1}{2}\x{\ell_1}{3}\x{\ell_1}{4}\x{\ell_1}{\ell_2}\x{\ell_2}{4}\x{\ell_2}{5}\x{\ell_2}{6}\x{\ell_2}{1}}.\hspace{-20pt}\label{omega_2_integral}}
This integral has been called $\Omega^{(2)}$ or $\widetilde{\Omega}^{(2)}$, depending on the choice of numerators $N_i$. In $x$-space, the factor $\x{\ell_1}{N_1}$ has the meaning of $(x_{\ell_1}\mi x_{N_1})^2$ where the point $x_{N_1}$ in (\ref{omega_2_integral}) would be defined as one of the two points light-like separated from all the points $\{x_1,x_2,x_3,x_4\}$. That is, $x_{N_1}$ is a solution to the system of quadratic equations
\eq{\x{N_1}{1}=\x{N_1}{2}=\x{N_1}{3}=\x{N_1}{4}=0\,.\label{defining_the_numerator_in_x_space}}
It is not hard to show that the solutions to (\ref{defining_the_numerator_in_x_space}) are not rationally related to the points $x_a$. We cannot resist mentioning here that the square root that is needed is nothing other than the familiar Gramian mentioned in \mbox{section \ref{subsec:dual_momenta_and_conformal_invariance}}:
\eq{\gramian{6}{\{123456\}}=\sqrt{(1\mi u_1\mi u_2\mi u_3)^2\mi 4 u_1u_2u_3}\,,\label{six_particle_gram_det}}
where the dual-conformal cross-ratios $u_i$ are defined as
\eq{\fwbox{0pt}{\hspace{-30pt}u_1\!\equiv\!\u{13}{46}\!=\!\frac{\x{1}{3}\x{4}{6}}{\x{1}{4}\x{3}{6}},\,\, u_2\!\equiv\!\u{24}{51}\!=\!\frac{\x{2}{4}\x{1}{5}}{\x{2}{5}\x{1}{4}},\,\, u_3\!\equiv\!\u{35}{62}\!=\!\frac{\x{3}{5}\x{2}{6}}{\x{3}{6}\x{2}{5}}.}\label{uvw_defined}}

In twistor space, by contrast, the equations (\ref{defining_the_numerator_in_x_space}) have a rather different meaning. The point $x_{N_1}$ would be associated with some line $(z_{N_1}^1,z_{N_1}^2)$ in twistor space---which we will sloppily denote `$(N_1)$'---that simultaneously intersects the four lines $\{(61),(12),(23),(34)\}$. Because two lines intersecting is equivalent to their combined span being less than full rank, this amounts to finding a solution to the equations
\eq{\ab{(N_1)\,61}=\ab{(N_1)\,12}=\ab{(N_1)\,23}=\ab{(N_1)\,34}=0.\label{defining_the_numerator_in_twistor_space}}
The two solutions to this system of equations are easy to find (especially if one thinks geometrically). They are
\eq{N_1=(13)\,,\qquad\overline{N_1}\equiv(612)\tncap(234)\,,\label{omega_2_numerators_in_twistor_space}}
where the notation used above was defined in \mbox{section \ref{subsec:momentum_twistors}} (see (\ref{definition_of_intersection_of_planes})).
(A more thorough discussion of the geometry here can be found in \mbox{ref.\ \cite{ArkaniHamed:2010gh}}.) It turns out that the points in $x$-space corresponding to these two solutions are complex conjugates if the momenta are real in $\mathbb{R}^{3,1}$ signature; more generally, they are related by parity. If both $N_i$ have the same chirality---for example, $\{N_1,N_2\}\!=\!\{(13),(46)\}$---then the integral is called $\Omega^{(2)}$; otherwise it is called $\widetilde{\Omega}^{(2)}$ (or its rotation by three). See \mbox{refs.\ \cite{Drummond:2010cz,omegaLadders}} for a more thorough discussion of these conventions. 

The point of the discussion above is that the tensor numerators $N_i$ defining the finite two-loop integrals (\ref{omega_2_integral}) can be defined rationally in twistor space, but not in $x$-space. This is of course related to the fact that {\it any} rational parameterization of momentum twistors will rationalize the Gramian determinant (\ref{six_particle_gram_det}). We review some of the more familiar parameterizations in the next subsection.

\paragraph{Feynman Parameterization of $\Omega^{(2)}$}~\\[-14pt]

Although we cannot express the integrand (\ref{omega_2_integral}) in $x$-space without introducing square roots (namely, the Gramian (\ref{six_particle_gram_det}) above), it turns out that $\Omega^{(2)}$ can be represented as a Feynman parameter integral rationally depending on dual-conformal cross-ratios:\footnote{This expression can be obtained as a restriction (and rotation of indices) of a more general result derived in \mbox{appendix \ref{appendix:derivation_of_the_octagon_ladder}}.}
\eq{\hspace{-30pt}\Omega^{(2)}(u_1,u_2,u_3)\equiv\!\!\int\limits_0^\infty\!\!d^5\!\vec{\alpha}\frac{u_3}{f_1f_2f_3f_4}\left(\frac{\alpha_4(1\mi u_1)}{f_2}\pl\frac{\alpha_5(1\mi u_2)\pl 1\mi u_1\mi u_2\pl u_3}{f_3}\mi1\right),\hspace{-20pt}\vspace{-10pt}\label{feynman_parameter_rep_of_omega_2}}
where
\vspace{-5pt}\eq{\begin{array}{l@{$\;\;\;\;\;$}l}f_1\equiv\alpha_1u_1\pl\alpha_1\alpha_2\pl \alpha_2\alpha_5 u_2\,,&f_2\equiv f_1\pl \alpha_2(\alpha_3\pl\alpha_4)\pl\alpha_3 u_1\pl\alpha_4(1\pl\alpha_5)\,,\\
f_3\equiv\alpha_1\pl\alpha_3\pl\alpha_5\pl u_3\,, &f_4\equiv\alpha_1\pl\alpha_3\pl\alpha_4\pl\alpha_5 u_2\,.
\end{array}\vspace{-5pt}\label{definition_of_factors_in_feynman_parameter_rep_of_omega_2}}
Notice that the form of (\ref{feynman_parameter_rep_of_omega_2}) makes it clear that the expression is single-valued over the space of positive cross-ratios, $u_i\!\in\!\mathbb{R}_+$---a space considerably larger than the Euclidean domain. One might therefore be optimistic that (\ref{feynman_parameter_rep_of_omega_2}) can be expressed in terms of iterated integrals depending rationally on the cross-ratios. However, this is not so---a fact that has long been known, and has led to a rich story of symbol alphabets (see e.g.\ \mbox{refs.\ \cite{Caron-Huot:2016owq,Drummond:2017ssj,Golden:2014xqa,Dixon:2011pw,Goncharov:2010jf,Golden:2013xva}}).  

From the point of view of direct integration via partial fractioning (as implemented in {\tt HyperInt}, for example), it is not hard to see that the square root related to the Gramian (\ref{six_particle_gram_det}) necessarily arises. (One way for the reader to confirm this would be to notice, for example, that all codimension-four residues associated with $f_i\!=\!0$ (in (\ref{definition_of_factors_in_feynman_parameter_rep_of_omega_2})) involve this square root.) Thus, if we wish to express hexagon functions in terms of iterated integrals, we should use variables that rationalize the root (\ref{six_particle_gram_det}).

\vspace{-2pt}
\subsubsection{Hexagon Kinematics in Momentum-Twistor Space}\label{subsubsec:hexagon_coordinates_survey}%
\vspace{-2pt}

Among the most familiar parameterizations of hexagon functions are the so-called $y$-variables \cite{Dixon:2011pw, Dixon:2011nj, Dixon:2012yy, Dixon:2013eka, Dixon:2014voa, Dixon:2014xca, Dixon:2014iba, Dixon:2015iva, Caron-Huot:2016owq, Dixon:2016apl}. As coordinates in the space of momentum twistors, they correspond to the functions
\eq{\hspace{-10pt}y_1\!\equiv\!\frac{\ab{4612}\ab{5123}\ab{3456}}{\ab{3451}\ab{4562}\ab{6123}},\; y_2\!\equiv\!\frac{\ab{2456}\ab{3561}\ab{1234}}{\ab{1235}\ab{2346}\ab{4561}},\; y_3\!\equiv\!\frac{\ab{2346}\ab{3451}\ab{5612}}{\ab{3561}\ab{4612}\ab{2345}}.\hspace{-10pt}\label{hexagon_y_variables_as_coordinates}}
They can also be thought of as {\it parametrizing} the kinematics through the matrix of momentum twistors
\eq{Z^{(6)}_{\vec{y}}\equiv\raisebox{-0pt}{$\left(\raisebox{28pt}{}\right.$}\begin{array}{@{}c@{$\;$}c@{$\;$}c@{$\;\;\;$}c@{$\;\;\;$}c@{$\;\;\;$}c@{}}1&0&0&1\mi y_1&1\mi y_1 y_3&0\\[-2pt]
0&1&1&y_1 y_2&y_3(1\pl y_1 y_2)\mi1&0\\[-2pt]
0&0&1&1&y_3&0\\[-2pt]
0&1&0&y_2\mi1&0&1\end{array}\raisebox{-0pt}{$\left.\raisebox{28pt}{}\right)$},\label{hexagon_y_variables_as_parameters_for_z}}
from which (\ref{hexagon_y_variables_as_coordinates}) can easily be confirmed. One advantage of thinking of the variables $y_i$ as {\it parameterizing} twistors rather than as `coordinates' (\ref{hexagon_y_variables_as_coordinates}) (namely, as maps from twistor space to $\mathbb{R}$) is that any other four-bracket can be easily computed directly from the matrix (\ref{hexagon_y_variables_as_parameters_for_z}). For example, it is easy to see that the $u_i$ defined in (\ref{uvw_defined}) are given by
\eq{u_1\!=\!\frac{y_1(1\mi y_2)(1\mi y_3)}{(1\mi y_1y_2)(1\mi y_3y_1)},\quad u_2\!=\!\frac{y_2(1\mi y_3)(1\mi y_1)}{(1\mi y_2y_3)(1\mi y_1y_2)},\quad u_3\!=\!\frac{y_3(1\mi y_1)(1\mi y_2)}{(1\mi y_3y_1)(1\mi y_2y_3)},\label{uvw_in_y_space}}
and that they rationalize the Gramian square root (\ref{six_particle_gram_det}). As such, integration by partial fractioning may proceed without unnecessary complications, and it is easy to re-express (\ref{feynman_parameter_rep_of_omega_2}) in terms of iterated integrals depending rationally on the $y_i$'s using standard techniques---e.g.\ those implemented in {\tt HyperInt}.

\paragraph{Choosing Good Charts}~\\[-14pt]

As we have mentioned, {\it any} rational parameterization of twistor space will rationalize the square root (\ref{six_particle_gram_det}), and thus allow for integration by partial fractioning. Therefore, we can search for parameterizations that satisfy further desirable criteria. For instance, the $y_i$ letters in (\ref{hexagon_y_variables_as_coordinates}) transform as $y_i\!\to\!1/y_{i+1}$ when the external particle indices are cycled, which is useful for expressing functions that respect dihedral symmetry. However, this is not necessarily desirable when expressing functions such as $\Omega^{(2)}$ that break this symmetry. That is, it is often better to choose a chart whose symmetries match those of the integral under study.

In additional to paying attention to symmetries, it is advantageous to choose charts that manifestly preserve the single-valuedness of Feynman integrals over the Euclidean domain. As described in \mbox{section \ref{subsec:momentum_twistors}}, this is easily done by choosing `positive' configurations of twistors; in particular, it is always possible to choose twistor-space parameterizations that map the positive domain to all of $\mathbb{R}_+$ (see \mbox{section \ref{subsec:momentum_twistors}}). One especially canonical example of such a chart is described in \mbox{appendix \ref{appendix:cluster_chart_comparisions}}. It corresponds to the edge variables associated with the boundary measurements of the following plabic graph:
\eq{\fwboxR{0pt}{}\fig{-29.67pt}{1}{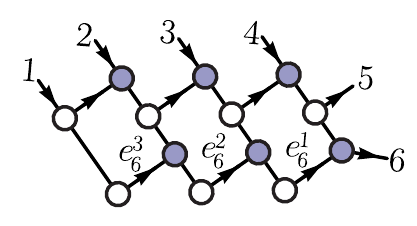}\bigger{\Leftrightarrow}\, Z_{\text{seed}}^{(6)}\equiv\!\raisebox{-0pt}{$\left(\raisebox{28pt}{}\right.$}\begin{array}{@{}c@{$\;\;$}c@{$\;\;$}c@{$\;\;$}c@{$\;\;$}c@{$\;\;$}c@{}}1&1\pl e_6^3&e_6^3&0&0&0\\[-2pt]0&1&1\pl e_6^2&e_6^2&0&0\\[-2pt]0&0&1&1\pl e_6^1&e_6^1&0\\[-2pt]0&0&0&1&1&1\end{array}\raisebox{-0pt}{$\left.\raisebox{28pt}{}\right)$}.\label{seed_plabic_graph_and_parameterization_for_g46}}
Viewed as a coordinate chart, these edge variables correspond to the cross-ratios
\eq{e_6^1\equiv\frac{\ab{1234}\ab{1256}}{\ab{1236}\ab{1245}},\; e_6^2\equiv\frac{\ab{1235}\ab{1456}}{\ab{1256}\ab{1345}},\; e_6^3\equiv\frac{\ab{1245}\ab{3456}}{\ab{1456}\ab{2345}}.\label{seed_chart_for_g46}}

However, for $\Omega^{(2)}$ defined in (\ref{omega_2_integral}) (and its `pentaladder' generalizations $\Omega^{(L)}$ \cite{Caron-Huot:2018dsv}) there seems to exist a superior chart---as measured by the simplicity of the iterated-integral expression when expressed in an appropriate basis. It originates from the boundary measurements of
\eq{\fwbox{100pt}{\fig{-29.67pt}{1}{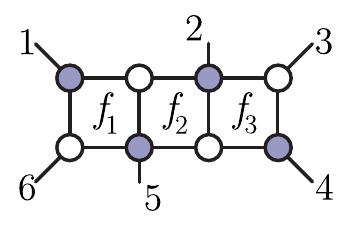}}\bigger{\Leftrightarrow}\;Z^{(6)}_{\Omega}\equiv\raisebox{-0pt}{$\left(\raisebox{28pt}{}\right.$}\begin{array}{@{}c@{$\;$}c@{$\;$}c@{$\;$}c@{$\;$}c@{$\;$}c@{}}1&0&0&1&1\pl f_2&0\\[-2pt]0&1&1&1&0&0\\[-2pt]0&0&1&1\pl f_3&f_3&0\\[-2pt]0&0&0&1&1\pl f_2(1\pl f_1)&1\end{array}\raisebox{-0pt}{$\left.\raisebox{28pt}{}\right)$},\label{optimal_parameterization_for_omega}}
where the coordinates can be thought to be defined as
\eq{f_1\equiv\frac{\ab{1346}\ab{2345}}{\ab{1234}\ab{3456}},\quad f_2\equiv\frac{\ab{1236}\ab{3456}}{\ab{1356}\ab{2346}},\quad f_3\equiv\frac{\ab{1256}\ab{1346}}{\ab{1236}\ab{1456}}.\label{optimal_chart_for_omega}}
In this chart, $\Omega^{(L)}$ can be expressed as a polynomial in $\log(f_2)$ and $\log (f_1 f_3)$ whose coefficients are (Goncharov) polylogarithms drawn from the set
\eq{\left\{ G_{\vec{w}}(f_2) \bigg| w_i\!\in\!\left\{ 0, \mi1, \frac{\mi1}{1\pl f_1}, \frac{\mi1}{1\pl f_3}, \frac{\mi1}{(1\pl f_1)(1\pl f_3)} \right\}\!\right\} \,}
that, furthermore, always take the form
\eq{G_{\fwboxL{27pt}{{\underbrace{0,\dots,0}_{L-1}}},\vec{w}'}(f_2)\,}
at $L$ loops. This chart for $\Omega^{(L)}$ was found by us through brute force surveys of edge and face charts associated with plabic graphs. But it is interesting to note that this chart was used (in the course of presenting a different set of coordinates) in \mbox{appendix A} of \mbox{ref.\ \cite{Basso:2013aha}}. 

Again, an advantage of considering $f_i$ letters as {\it parameters} as opposed to {\it coordinates} is that we can easily compute other functions in terms of the $f_i$ letters by simply evaluating determinants on the twistors given in (\ref{optimal_parameterization_for_omega}); thus, the cross-ratios $u_i$ would be parameterized in this chart by
\eq{u_1\!=\!\frac{f_2}{1\pl f_2}\,,\quad u_2\!=\!\frac{1}{1\pl (1\pl f_1)f_2(1\pl f_3)}\,,\quad u_3\!=\!\frac{f_1f_2f_3}{(1\pl f_2)(1\pl (1\pl f_1)f_2(1\pl f_3))}\,.\label{uvw_in_omega_optimal_chart}}

The task of tailoring a chart to a specific integral in this fashion, rendering the final expression maximally simple, deserves more attention. For now though, we leave this to future work and focus on charts with more general advantages.

\subsection{Heptagon Integrals}\label{subsec:heptagons}%
\vspace{-2pt}

\paragraph{Dual-Conformal Kinematics for Seven Particles}~\\[-14pt]

As reviewed in \mbox{section \ref{subsec:dual_momenta_and_conformal_invariance}}, a six-dimensional space of conformally invariant cross-ratios can be formed out of seven light-like separated momenta. It is simple to convince oneself that this is fewer than the number of multiplicatively independent cross-ratios by constructing a cross-ratio that involves only a single two-particle invariant, as this invariant will generate a complete seven-orbit of multiplicatively independent cross-ratios under the dihedral group. The redundancy in these variables follows from the fact that only six cross-ratios can be independent, which also implies that the Gramian determinant (\ref{gramian_identities}) must vanish. As it turns out, the single such constraint at seven points is quadratic in each cross-ratio---and thus, despite the fact that Feynman-parametric representations of most (if not all) loop integrals can be found which depend exclusively (and perhaps even rationally) on dual-conformal cross-ratios, eliminating any one of them will introduce a (fairly complicated) square root into our description. As in the case of six particles, however, the arguments of these square roots are recognizable as the $6\!\times\!6$ Gramian determinants $\Delta_7^A$, of which there are $\binom{7}{6} = 7$. Therefore, as observed in section~\ref{sec:kinematics}, these roots will be rationalized by twistors. 

\paragraph{Heptagon Functions at Two Loops}~\\[-14pt]

There are three classes of double-pentagon integral topologies that we can define for seven particles. At two loops, they correspond to 
\eq{\begin{array}{@{}c@{}}~\\[-15pt]\fig{-29.67pt}{1}{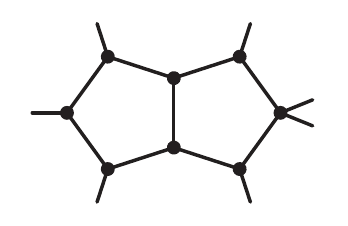}\\[-15pt]A\end{array},\begin{array}{@{}c@{}}~\\[-15pt]\fig{-29.67pt}{1}{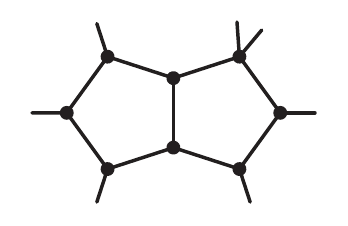}\\[-15pt]B\end{array},\begin{array}{@{}c@{}}~\\[-15pt]\fig{-29.67pt}{1}{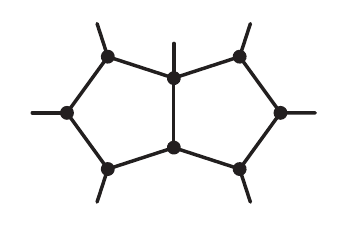}\\[-15pt]C\end{array}.\label{two_loop_seven_point_topologies}}
For each of these sets of propagators, there is some number of loop-dependent numerators which render the integrals infrared finite, pure, and dual-conformally invariant. Different choices for these numerators give rise to different integrals. It turns out that only the first and last of the topologies in (\ref{two_loop_seven_point_topologies}) have cuts with non-vanishing support for MHV amplitudes; and for each of these, infrared finiteness uniquely dictates their possible numerators to be one of four possibilities. As we will soon describe, there are more choices of numerators for the second topology. More generally, we refer interested readers to \mbox{ref.\ \cite{Bourjaily:2015jna}} (see also ref.\ \cite{Bourjaily:2017wjl}) for conventional definitions of such numerators and the logic behind the possible choices (as these affect the representation of amplitudes). 

From the discussion in \mbox{section \ref{subsec:dual_momenta_and_conformal_invariance}}, we see that the first two integrals in (\ref{two_loop_seven_point_topologies}) depend on four DCI parameters each, as only five of the six dual points on which they depend are light-like separated. Parametrizing each of these integrals in terms of the right number of variables in twistor space (by going to appropriate boundaries of the positive Grassmannian), they prove to be only mildly more difficult than their six-point counterparts. The last integral, however, depends on all six independent cross-ratios for massless seven-particle kinematics, and represents a harder class of Feynman integrals. Let us now discuss each in turn.

\subsubsection{Heptagon A: An MHV Integral Topology (and its Ladder)}\label{subsubsec:heptagon_a_section}

Let us first consider the first topology of (\ref{two_loop_seven_point_topologies})
\eq{\fwbox{0pt}{\fig{-29.67pt}{1}{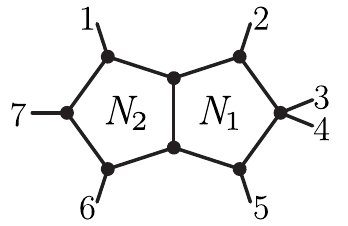}\bigger{\Leftrightarrow}\fig{-29.67pt}{1}{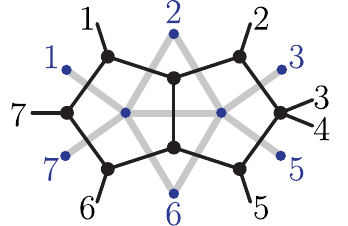}\bigger{\Leftrightarrow}\fig{-29.67pt}{1}{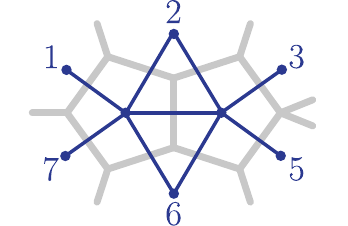}\fwboxL{0pt}{,}} \label{heptagon_a_dual_sequence_figures}}
whose dual graph we have also drawn. It corresponds to the integral
\eq{\hspace{-20pt}I_{7,A}^{(2)}\equiv\!\!\int\!\!d^4\ell_1\,d^4\ell_2\frac{\x{\ell_1}{N_1}\x{2}{6}\x{\ell_2}{N_2}}{\x{\ell_1}{2}\x{\ell_1}{3}\x{\ell_1}{5}\x{\ell_1}{6}\x{\ell_1}{\ell_2}\x{\ell_2}{6}\x{\ell_2}{7}\x{\ell_2}{1}\x{\ell_2}{2}}\,.\hspace{-20pt}\label{two_loop_heptagon_a_loop_space_definition}}
We choose the numerators to be
\eq{N_1\!\Leftrightarrow\! (123)\tcap(456)\,,\qquad N_2\!\Leftrightarrow\!(567)\tcap(712)\,,\label{heptagon_a_ladder_numerators}}
so as to focus on the version of this integral that contributes to the MHV amplitude. (This numerator also corresponds to the $z_7\!\parallel\!z_8$ limit of~\eqref{octa_ladder_numerators_appendix}.) (There is of course a choice of numerators analogous to those of the hexagon $\widetilde{\Omega}^{(2)}$ integrals. As far as integration is concerned, there is no substantive difference in difficulty relative to the choice relevant to MHV integrals.) We hereafter refer to this integral (and its $L$-loop generalization, given below) as `heptagon A'. 

Since heptagon A does not depend on ${\color{hred}x_4}$, it can be thought of as living entirely in the codimension-two positroid cell defined by
\eq{\ab{12\,{\color{hred}34}}= \ab{{\color{hred}34}\,56}= 0\,. \label{heptagon_a_boundary}}
These invariants can each be isolated into a single cross-ratio, both of which will then drop out of this limit. This leaves four independent cross-ratios, which we can take to be
\eq{\hspace{-85pt}u_1\equiv\u{13}{57}\,,\quad u_2\equiv\u{25}{61}\,,\quad u_3\equiv\u{36}{72}\,,\quad u_4\equiv\u{35}{62}\,.\hspace{-50pt}\label{heptagon_a_specific_cross_ratios}}
The positroid associated with the conditions (\ref{heptagon_a_boundary}) can be parameterized by the boundary measurements associated with the plabic graph 
\eq{\begin{array}{@{}c@{}}\fig{-29.67pt}{1}{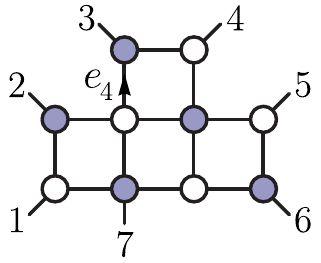}\\[-2pt]\fwboxR{0pt}{\sigma\!\equiv}\{{\color{black}4},7,{\color{black}6},8,9,10,12\}\end{array}\bigger{\Rightarrow} \ \ Z^{(7)}_{\text{A}}\equiv\raisebox{-0pt}{$\left(\raisebox{28pt}{}\right.$}\begin{array}{@{}c@{$\;$}c@{$\;$}c@{$\;$}c@{$\;$}c@{$\;$}c@{$\;$}c@{}}1&0&0&e_4&0&\mi\!1&\mi\!1\mi e_2(1\pl e_3)\\[-2pt]1&1&0&0&0&0&\mi e_2e_3\\[-2pt]0&0&1&1\pl e_1e_4&1&1&0\\[-2pt]0&0&0&0&1&1\pl e_1&e_1\end{array}\raisebox{-0pt}{$\left.\raisebox{28pt}{}\right).$}\label{heptagon_a_bespoke_chart_with_boundary_measurements}}
Viewed as coordinates (on the space of momentum twistors), these correspond to the chart
\eq{\begin{split}e_1\!\equiv\!\frac{\ab{1237}\ab{1256}}{\ab{1235}\ab{1267}},\, e_2\!\equiv\!\frac{\ab{1235}\ab{1567}\ab{2456}}{\ab{1257}\ab{1456}\ab{2356}},\, e_3\!\equiv\!\frac{\ab{1256}\ab{4567}}{\ab{1567}\ab{2456}},\,e_4\!\equiv\!\frac{\ab{1267}\ab{2345}}{\ab{1237}\ab{2456}}\end{split}.\label{heptagon_a_bespoke_coordinates}}
It is worth highlighting some of the desirable features that are built into this chart. In addition to parametrizing heptagon A in terms of the correct number of variables, this parametrization smoothly degenerates to the preferred chart for the hexagon integral described in the last subsection. This behavior is encoded in the variable $e_4$, which is associated with a specific edge in the plabic graph---as highlighted in~\eqref{heptagon_a_bespoke_chart_with_boundary_measurements}. Namely, sending $e_4\!\to\!0$ corresponds to deleting this edge, which has the effect of reducing
\eq{\begin{array}{@{}c@{}}\fig{-29.67pt}{1}{heptagon_a_plabic_graph}\\[-2pt]\fwboxR{0pt}{\sigma\!\equiv}\{{\color{hred}4},7,{\color{hred}6},8,9,10,12\}\end{array}\underset{e_4\to0}{\longrightarrow}\begin{array}{@{}c@{}}\fig{-29.67pt}{1}{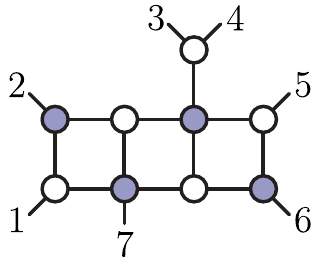}\\[-2pt]\{{\color{hred}6},7,{\color{hred}4},8,9,10,12\}\fwboxL{0pt}{\equiv\!\sigma'\!=\!(13)\!\circ\!\sigma}\end{array}\,.\label{heptagon_a_bespoke_chart_with_boundaries_and_permutation_labels}}
(Recall from \mbox{section \ref{subsec:momentum_twistors}} that two legs attached to same white vertex correspond, geometrically, to twistors that are proportional to one another.) Thus, the plabic graph on the right above exactly matches~\eqref{optimal_parameterization_for_omega} (upon relabeling). More concretely, $e_i$ of the chart (\ref{heptagon_a_bespoke_chart_with_boundaries_and_permutation_labels}) matches $f_i$ of (\ref{optimal_parameterization_for_omega}) upon sending $e_4\!\to\!0$ and relabeling the twistors $\{z_1,\ldots,z_7\}$ to be $\{z_3,z_4,z_5,z_5,z_6,z_1,z_2\}$.

(The attentive reader will see that we can easily iterate this type of construction, embedding the chart in~\eqref{heptagon_a_bespoke_chart_with_boundary_measurements} into the parametrization of any eight-point integral that includes heptagon A as a limit. It is hoped that embedding lower-point charts (that permit especially parsimonious representations) in this way will give rise to similar simplifications at higher points, since the problem of predicting a good parametrization and function basis directly from integrands remains largely unexplored.)

In terms of the parameterization~\eqref{heptagon_a_bespoke_chart_with_boundary_measurements}, the cross-ratios~\eqref{heptagon_a_specific_cross_ratios} become
\eq{\begin{array}{@{}l@{$\;\;\;\;$}l@{}}\displaystyle u_1\!=\!\frac{e_1e_2e_3}{(1\pl(1\pl e_1)e_2(1\pl e_3))(1\pl(1\pl e_1 e_4)e_2)}\,,&\displaystyle u_2\!=\!\frac{(1\pl e_1 e_4)e_2}{1\pl(1\pl e_1e_4)e_2}\,,\\[10pt]
\displaystyle u_3\!=\!\frac{1}{1\pl(1\pl e_1)e_2(1\pl e_3)}\,,&\displaystyle u_4\!=\!\frac{e_1e_4}{1\pl e_1e_4}\,.\end{array}\label{heptagon_a_us_in_es}}
As highlighted in previous sections, all of the square roots in cross-ratio space that result from the Gramian determinant constraint are rationalized in these variables. In fact, only one such square root appears in heptagon A: 
\eq{\begin{split}\gramian{7}{\{123567\}}&=\sqrt{(1\mi u_1\mi u_2\mi u_3\pl u_2u_3u_4)^2\mi 4 u_1u_2u_3(1\mi u_4)}\\&=\frac{e_2(e_1\mi e_3)}{(1\pl(1\pl e_1)e_2(1\pl e_3))(1\pl e_2(1\pl e_1 e_4))}\,.\end{split}\label{heptagon_a_gramian_in_es}}
This is simply due to the fact that all other seven-point square roots of this type depend on $x_4$. In these variables, the integration of~\eqref{two_loop_heptagon_a_loop_space_definition} thus becomes algorithmic, and can be carried out in {\tt HyperInt} in tens of seconds.

All of the above analysis extends unproblematically to the higher-loop analogs of heptagon A, in which a ladder of boxes is added between the pentagons:
\eq{\fig{-29.67pt}{1}{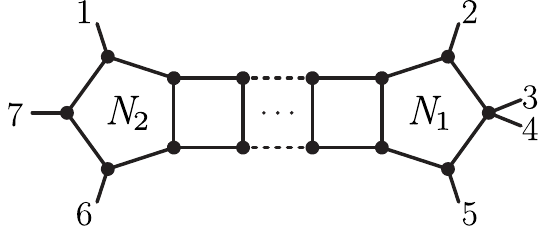}\,.\label{heptagon_a_ladder_image}} 
Following the logic of \mbox{appendix \ref{appendix:derivation_of_the_octagon_ladder}}, this integral can be put in the form
\eq{\hspace{-20pt}I_{7,A}^{(L)}\equiv\!\!\int\!\!d^{4L}\vec{\ell}\frac{\x{\ell_1}{N_1}\x{2}{6}^{L-1}\x{\ell_L}{N_2}}{\x{\ell_1}{3}\x{\ell_1}{5}\big(\prod_{i=1}^L\x{\ell_i}{2}\x{\ell_i}{6}\big)\big(\x{\ell_1}{\ell_2}\cdots\x{\ell_{L-1}}{\ell_L}\big)\x{\ell_L}{7}\x{\ell_L}{1}}\,.\!\hspace{-20pt}\label{heptagon_a_ladder_integral_loop_space_definition}}
(Alternately, we can take the $z_7\!\parallel\!z_8$ limit of~\eqref{octa_ladder_ultimate_form}). The only barriers to carrying out these integrations are limits on time and memory, and we have evaluated both the three- and four-loop integrals. We find they can be expressed as polynomials in $\log (e_1)$ and $\log(e_3)$ whose coefficients are polylogarithms drawn from the basis
\eq{\begin{split}\bigg\{ G_{\vec{w}}(e_2) \bigg| w_i\!\in\!&\left\{  0, \mi1, \frac{\mi1}{1\pl e_1}, \frac{\mi1}{1\pl e_3}, \frac{\mi1}{(1\pl e_1)(1\pl e_3)} \right\}\!,\\
 G_{\vec{w}}(e_4) \bigg| w_i\!\in\!&\,\bigg\{  0, \mi\frac{1}{e_1}, \mi\frac{1\pl e_2}{e_1 e_2}, \mi\frac{1\pl e_2(1\pl e_1)}{e_1 e_2 (1\pl e_1)}, \mi\frac{1\pl e_2(1\pl e_3)}{e_1 e_2 (1\pl e_3)},\\ 
& \hphantom{\,\bigg\{}\mi\frac{1\pl e_2(1\pl e_1)(1\pl e_3)}{e_1 e_2 (1\pl e_1)(1\pl e_3)}, \mi\frac{(1\pl e_2(1\pl e_1))(1\pl e_2(1\pl e_3))}{e_1 e_2 (1\pl e_2 (1\pl e_1)(1\pl e_3))} \bigg\}\!\bigg\} \,.\end{split}}

In this basis, taking the $e_4\!\to\!0$ limit is remarkably simple, as it just entails setting all $G_{\vec{w}}(e_4)$ equal to zero (no logarithmic singularities in $e_4$ survive this procedure). Taking this limit, the functions we obtain can be immediately identified as $\Omega^{(L)}$ in the appropriate chart.

We include just the two- and three-loop results with the submission in ancillary files, as the four-loop result is too large to include.

\subsubsection{Heptagon B: A non-MHV Integral Topology (and its Ladder)}\label{subsubsec:heptagon_b_section}
 
Let us now turn to the second class of heptagon integral topologies in (\ref{two_loop_seven_point_topologies}), which we shall refer to as the `heptagon B' class:
\eq{\fwbox{0pt}{\fig{-29.67pt}{1}{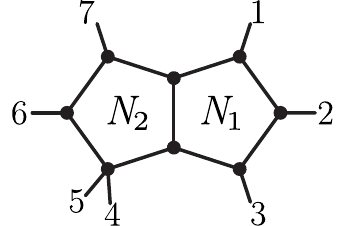}\bigger{\Leftrightarrow}\fig{-29.67pt}{1}{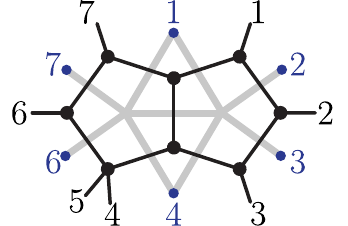}\bigger{\Leftrightarrow}\fig{-29.67pt}{1}{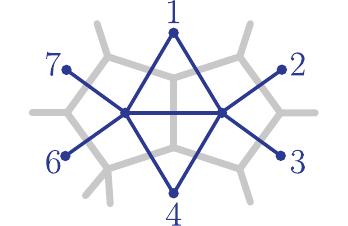}\fwboxL{0pt}{.}}\label{heptagon_b_dual_sequence_figures}}
Unlike the heptagon A integrals, infrared finiteness by itself is insufficient to uniquely determine the possible numerators of (\ref{heptagon_b_dual_sequence_figures})---at least not beyond a one-parameter ambiguity. The easiest way to see this is to notice that this integral topology admits a pentabox sub-topology,
\eq{\fig{-29.67pt}{1}{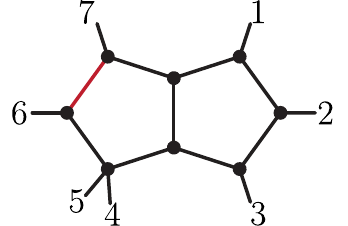} \quad \bigger{\supset} \quad \fig{-29.67pt}{1}{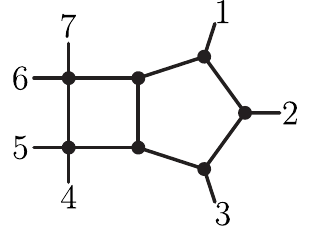}\,,\label{heptagon_b_contact_term_figures}}
for which (two) numerators exist that render it infrared finite, pure, and dual-conformally invariant \cite{Drummond:2010cz}. Thus, there is at least one continuous family of potential numerators for heptagon B---differing by an arbitrary contribution from these pentabox contact-terms:
\eq{\x{\ell_2}{N_2}\to\Big(\x{\ell_2}{N_2}\pl\delta\x{\ell_2}{7}\Big)\,.\label{heptagon_b_delta_ambiguity}}
It is worth mentioning that no choice of $\delta$ in (\ref{heptagon_b_delta_ambiguity}) can eliminate all the residues of heptagon B which cut the propagators of the pentabox. Because of this, the specific choice made for the numerator should be informed by the representation of amplitudes---for example, by which pentabox integrals (and which cuts are chosen to normalize them) are chosen for an integrand basis.

In the present work, we have chosen to adopt the four heptagon B numerators defined in \mbox{ref.\ \cite{Bourjaily:2015jna}}, which we describe in more detail in \mbox{appendix \ref{appendix:nonmhv_numerator_details}}. One consequence of following these conventions is that all four of our heptagon B integrals include pentaboxes as sub-{\it integrals} as in (\ref{heptagon_b_contact_term_figures}). This means that these integrals do not smoothly degenerate to $\Omega^{(2)}$, as
\eq{ \fig{-29.67pt}{1}{heptagon_b_integral_contact_term}\;\;\underset{z_7\!\parallel\! z_6}{\bigger{\longrightarrow}}\;\;\fig{-29.67pt}{1}{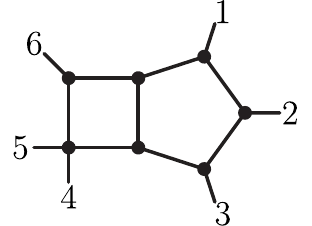}\,,\label{heptagon_b_contact_term_figures_2}}
which clearly diverges. (There do exist choices of numerators for heptagon B that eliminate support on these sub-integrals, but the change would have a considerable effect on how amplitudes would be represented compared to the basis of integrals in ref.~\cite{Bourjaily:2015jna}.) Again, we refer the reader to \mbox{appendix \ref{appendix:nonmhv_numerator_details}} for the precise definition of (the four numerators of) heptagon B. 

As with heptagon A, there is a natural $L$-loop ladder generalization of \mbox{heptagon B}:~\\[-11pt]
\eq{\fig{-29.67pt}{1}{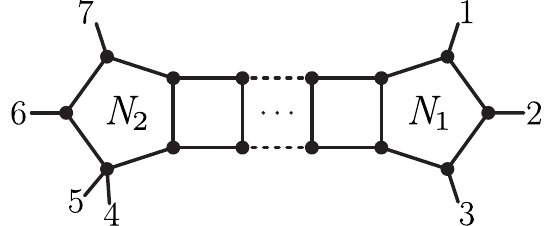}\,.\label{heptagon_b_ladder_image}} 
Also like heptagon A, this class of integrals does not depend on the entire space of seven-particle kinematics. Because heptagon B in  (\ref{heptagon_b_dual_sequence_figures}) is independent of the point $x_5$, we have the freedom to eliminate the dependence on the line $({\color{hred}45})$ in twistor space by imposing the constraints
\eq{\ab{23\,{\color{hred}45}} = \ab{{\color{hred}45}\,67} = 0\,.\label{heptagon_b_boundary}}
The codimension-two positroid configuration obtained by the constraints (\ref{heptagon_b_boundary}) is labeled by the permutation whose canonical (meaning, lexicographic bridge-constructed) plabic graph representative is given by
\eq{\sigma\!=\!\{6,5,8,7,9,10,11\}\,\,\bigger{\Rightarrow}\,\,\Gamma_\sigma\equiv\hspace{-10pt}\fig{-42.185pt}{1}{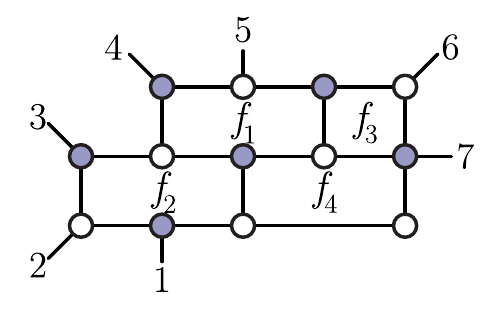}\,.\label{heptagon_b_plabic_graph}}
The boundary measurements of this plabic graph representative parameterize the momentum twistors according to
\eq{Z^{(7)}_{\text{B}}\equiv\raisebox{-0pt}{$\left(\raisebox{28pt}{}\right.$}\begin{array}{@{}c@{$\;$}c@{$\;$}c@{$\;$}c@{$\;$}c@{$\;$}c@{$\;$}c@{}}1&1&1&0&0&0&0\\[-2pt]0&1&1\pl f_2(1\pl f_4)&f_2(1\pl(1\pl f_1)f_4)&f_1f_2f_4&0&0\\[-2pt]0&0&1&1\pl f_1(1\pl f_3)&f_1(1\pl f_3)&f_1f_3&0\\[-2pt]0&0&0&1&1&1&1\end{array}\raisebox{-0pt}{$\left.\raisebox{28pt}{}\right),$}\label{heptagon_b_boundary_measurements}}
which, viewed as coordinates, correspond to the chart
\eq{\begin{split}f_1\!\equiv\!\frac{\ab{1256}\ab{1567}\ab{2346}}{\ab{1236}\ab{1456}\ab{2567}},\, f_2\!\equiv\!\frac{\ab{1567}\ab{2356}}{\ab{1256}\ab{3567}},\, f_3\!\equiv\!\frac{\ab{1235}\ab{1267}}{\ab{1237}\ab{1256}},\, f_4\!\equiv\!\frac{\ab{1236}\ab{2567}}{\ab{1267}\ab{2356}}\end{split}.\label{heptagon_b_bespoke_coordinates}}
Parameterized in this way, we find no obstruction to direct integration (besides memory and time). We have carried out integrations for each of the four choices of numerators outlined in \mbox{appendix \ref{appendix:nonmhv_numerator_details}}. (Only three of these choices of numerators give different results upon integration.) We find that they can all be expressed as polynomials in $\log (f_2)$ whose coefficients are polylogarithms drawn from the basis
\begin{align}\bigg\{ G_{\vec{w}}(f_1) \bigg| w_i\!\in\!&\left\{0, \frac{\mi1}{1\pl f_2} \right\}\! ,\nonumber\\G_{\vec{w}}(f_4) \bigg| w_i\!\in\!&\left\{0, \mi 1,\frac{\mi1}{1\pl f_1},\mi\frac{1\pl f_2}{f_2} \right\} \!,\\G_{\vec{w}}(f_3) \bigg| w_i\!\in\!&\left\{0, \mi 1,\frac{\mi1}{1\pl f_4}, \mi\frac{1\pl f_1}{f_1}, \mi\frac{1\pl f_2}{1\pl f_2(1\pl f_4)}, \mi\frac{1\pl f_1(1\pl f_2)}{f_1 (1\pl f_2)},\mi\frac{1\pl f_1(1\pl f_2)}{f_1(1\pl f_2(1\pl f_4))} \right\}\!\!\!\bigg\} .\nonumber\end{align}
Even at four loops, these can be integrated in less than a day. As in the case of heptagon A, we include two-loop and three-loop expressions for all heptagon B integrals as ancillary files to this work's submission to the {\tt arXiv}.

\subsubsection{Heptagon C: An Algorithmic Obstruction?}\label{subsubsec:heptagon_c_section}

The last heptagon integral topology to consider at two loops is `heptagon C':
\eq{\fwbox{0pt}{\fig{-29.67pt}{1}{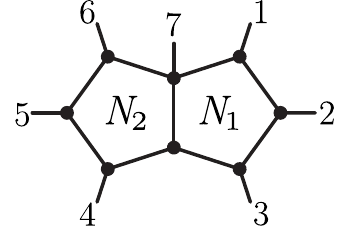}\bigger{\Leftrightarrow}\fig{-29.67pt}{1}{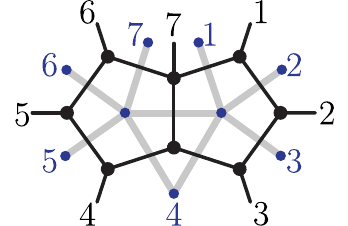}\bigger{\Leftrightarrow}\fig{-29.67pt}{1}{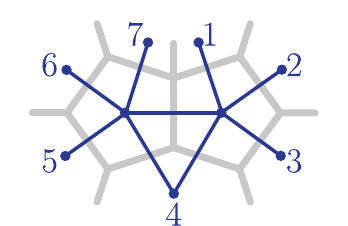}\fwbox{0pt}{.}}\label{heptagon_c_dual_sequence_figures}}
There is no unique way to generalize this topology to higher loops; as such, we consider only the two-loop topology above. 

In contrast to the other heptagon integrals, heptagon C depends on all six independent heptagon cross-ratios. For the sake of concreteness, we may choose to parameterize these degrees of freedom according to the canonical `seed' chart associated with the top cell of $G_+(4,7)$ as given in appendix~\ref{appendix:cluster_chart_comparisions}:
\eq{\fig{-28.185pt}{1}{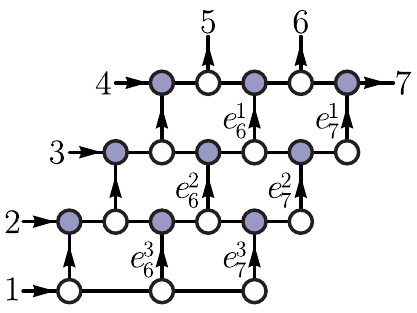}\hspace{-25pt}\bigger{\Rightarrow}Z^{(7)}_{\text{seed}}\!\equiv\!\raisebox{-0pt}{$\left(\raisebox{28pt}{}\right.$}\begin{array}{@{}c@{$\;\;$}c@{$\;\;$}c@{$\;\;$}c@{$\;\;$}c@{$\;\;$}c@{$\;\;$}c@{}}1&1\pl e_6^3\pl e_7^3&e_6^3\pl (1\pl e_6^2)e_7^3&e_6^2e_7^3&0&0&0\\[-2pt]0&1&1\pl e_6^2\pl e_7^2&e_6^2\pl(1\pl e_6^1)e_7^2&e_6^1e_7^2&0&0\\[-2pt]0&0&1&1\pl e_6^1\pl e_7^1&e_6^1\pl e_7^1&e_7^1&0\\[-2pt]0&0&0&1&1&1&1\end{array}\raisebox{-0pt}{$\left.\raisebox{28pt}{}\right)$}.\label{heptagon_seed_chart_with_boundary_measurements}}
Viewed as coordinates, the edge variables correspond to the chart 
\vspace{-32pt}\eq{\fwbox{0pt}{\begin{array}{@{}l@{$\;$}l@{$\;$}l@{}}~\\[14pt]\displaystyle e_6^1\equiv\frac{\ab{1234}\ab{1256}}{\ab{1236}\ab{1245}},\,&\displaystyle e_6^2\equiv\frac{\ab{1235}\ab{1456}}{\ab{1256}\ab{1345}},\,&\displaystyle e_6^3\equiv\frac{\ab{1245}\ab{3456}}{\ab{1456}\ab{2345}},\\
\displaystyle e_7^1\equiv\frac{\ab{1234}\ab{1235}\ab{1267}}{\ab{1236}\ab{1237}\ab{1245}},\,&\displaystyle e_7^2\equiv\frac{\ab{1236}\ab{1245}\ab{1567}}{\ab{1256}\ab{1267}\ab{1345}},\,&\displaystyle e_7^3\equiv\frac{\ab{1256}\ab{1345}\ab{4567}}{\ab{1456}\ab{1567}\ab{2345}}.\end{array}}\label{seven_point_seed_coordinate_chart}}

Like heptagon A, heptagon C is finite in the limit of $p_7\!\to\!0$ to $\Omega^{(2)}$. This is true already at the integrand level. (Notice that, unlike heptagon B, heptagon C does not have any finite pentabox sub-topologies.) Setting the {\it momentum} of particle $7$ to zero is a codimension-three constraint. In terms of the chart (\ref{heptagon_seed_chart_with_boundary_measurements}), this can be realized by sending the edge variables $e_7^1$, $e_7^2$ and $e_7^3$ to zero:
\eq{\fig{-36.185pt}{1}{heptagon_seed_plabic_graph}\hspace{-5pt}\underset{e_7^i\to0}{\longrightarrow}\hspace{5pt}\fig{-36.185pt}{1}{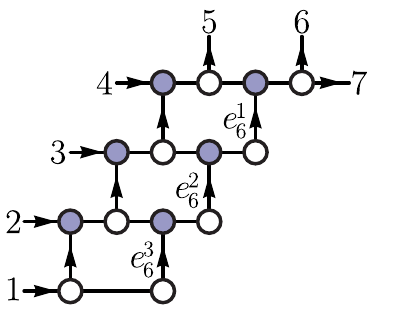}\,.\label{seed_chart_boundary_to_omega}}

As discussed above, one huge advantage of using a momentum-twistor parameterization such as (\ref{heptagon_seed_chart_with_boundary_measurements}) as opposed to cross-ratios is that it rationalizes all the square roots associated with $6\!\times\!6$ Gramian determinants. For example, 
\eq{\gramian{7}{\{123456\}}=\frac{e_6^2(1\mi e_6^3(e_6^1\pl e_7^1))}{(1\pl e_6^2(1\pl e_6^3))(1\pl e_6^1(1\pl e_6^2)\pl e_7^1(1\pl e_6^2\pl e_7^2))}\,.\label{heptagon_gramian_123456_in_seed_chart}}
(But of course, expressed in terms of cross-ratios, heptagon C would involve all seven such Gramian determinants---not just the one above.)

The most important novelty about heptagon C, however, is not the kinematic dependence of the integral. Rather, it turns out that one cannot (na\"{i}vely) integrate heptagon C via partial fractions as we did for the other examples without encountering an obstruction. Namely, after three integrations (of a five-fold Feynman parameter representation of the integral), one arrives at irreducible quadratic divisors in both of the remaining integration variables. Partial fractioning one of these would introduce a square root that depends on the final integration variables (which itself obstructs straight-forward integration). We discussed a similar obstacle in \mbox{section \ref{subsec:partial_fractioning_roots}}; in this case (and in the notation of \eqref{eq: partial fractioning}) the square root $\sqrt{f^2\mi g}$ would now depend on the last integration variable.

This type of obstacle to linear reducibility has been encountered previously in the literature, and is discussed in some illustrative examples in \mbox{ref.\ \cite{Panzer:2014gra}}. We emphasize that linear reducibility is a chart-dependent statement rather than a statement about the intrinsic properties of the integral itself. One way forward is to change integration variables (not the kinematic variables) to rationalize these square roots, making the integral linearly reducible---though it is not known how this can be done in general cases. In the present case, it turns out that even this can be avoided, as we describe in a forthcoming work \cite{us_in_the_future}.

\newpage\vspace{-2pt}
\subsection{Octagon Integrals (and Beyond)}\label{subsec:octagons_and_beyond}%
\vspace{-2pt}

At eight points, a much larger set of topologies contribute to two-loop amplitudes. We will concern ourselves with three representative examples:
\vspace{-10pt}\eq{\begin{array}{@{}c@{}}~\\[-12pt]\fig{-29.67pt}{1}{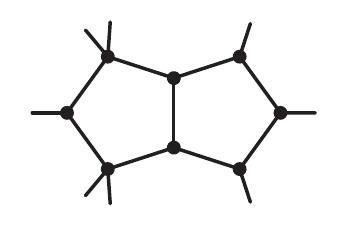}\\[-12pt]A\end{array},\begin{array}{@{}c@{}}~\\[-12pt]\fig{-29.67pt}{1}{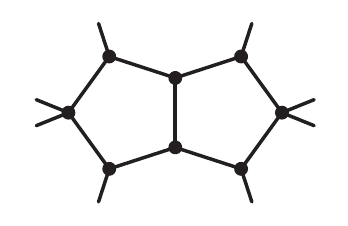}\\[-12pt]B\end{array},\cdots,\begin{array}{@{}c@{}}~\\[-12pt]\fig{-29.67pt}{1}{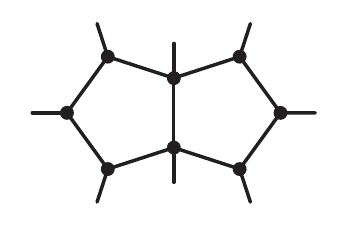}\\[-12pt]K\end{array}.\label{two_loop_eight_point_topologies}\vspace{-5pt}}
At one end of the spectrum, we have diagrams that depend on only five variables (six dual points, four pairs of which are light-like separated); on the other end, we have a diagram `K' that depends on the full set of eight dual points---nine independent cross-ratios---that represents the hardest case for our methods (although not because of the number of degrees of freedom).

\vspace{-2pt}
\subsubsection{Octagon A: An Eight-Point Integral Through Four Loops}\label{subsubsec:octagon_two_loop_discussion}%
\vspace{-2pt}

Let us start with the first octagon in \eqref{two_loop_eight_point_topologies}, which is a direct generalization of heptagon B:
\eq{\fwbox{0pt}{\fig{-29.67pt}{1}{octagon_a_integral}\bigger{\Leftrightarrow}\fig{-29.67pt}{1}{octagon_a_integral_with_dual}\bigger{\Leftrightarrow}
\fig{-29.67pt}{1}{octagon_a_integral_dual}\fwboxL{0pt}{.}}\label{octagon_a_dual_sequence}}
(We apologize to the reader that this integral is called octagon {\it A}.) Like heptagon B, the four pairs of numerators we take to define octagon A are described in detail in \mbox{appendix \ref{appendix:nonmhv_numerator_details}}. Similar to what we have seen before, the presence of (finite) pentabox sub-{\it integrals} in octagon A's definition prevents any smooth degeneration to lower multiplicity (as at least one of these pentaboxes will diverge in any degeneration). Thus, we need not concern ourselves with finding a coordinate chart that admits a smooth degeneration to e.g.\ heptagon B. 

The five independent degrees of freedom which parameterize octagon A can be obtained from the nine describing a generic configuration of eight momentum twistors by imposing the boundaries
\eq{\ab{23\,{\color{hred}45}} = \ab{{\color{hred}45}\,67} =\ab{56\,{\color{hred}78}}=\ab{{\color{hred}78}\,12}=0\,.\label{octagon_a_boundary}}
This corresponds to a positroid labeled by the permutation whose canonical (meaning, lexicographic bridge-constructed) plabic graph representative is
\vspace{-4pt}\eq{\sigma\!=\!\{6,5,9,7,8,11,10,12\}\,\,\bigger{\Rightarrow}\,\,\Gamma_\sigma\equiv\hspace{-10pt}\fig{-42.185pt}{1}{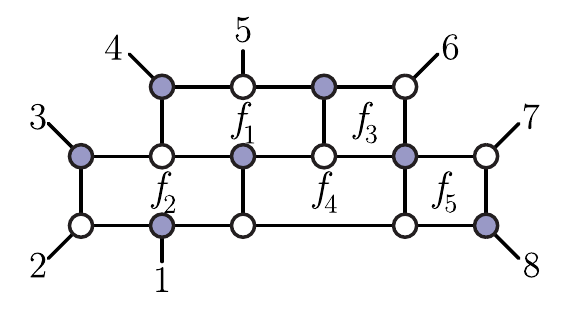}.\label{octagon_a_plabic_graph}}
The boundary measurements of this plabic graph can be understood to parameterize the momentum twistors, resulting in
\eq{Z^{(8)}_{\text{A}}\equiv\raisebox{-0pt}{$\left(\raisebox{28pt}{}\right.$}\begin{array}{@{}c@{$\;$}c@{$\;$}c@{$\;$}c@{$\;$}c@{$\;$}c@{$\;$}c@{$\;$}c@{}}1&1&1&0&0&0&0&0\\[-2pt]0&1&1\pl f_2(1\pl f_4)&f_2(1\pl(1\pl f_1)f_4)&f_1f_2f_4&0&0&f_1f_2f_3f_4f_5\\[-2pt]0&0&1&1\pl f_1(1\pl f_3)&f_1(1\pl f_3)&f_1f_3&0&0\\[-2pt]0&0&0&1&1&1&1&1\end{array}\raisebox{-0pt}{$\left.\raisebox{28pt}{}\right).$}\label{octagon_a_boundary_measurements}}
As before, we can also consider the face variables as {\it coordinates} (as opposed to {\it parameters}) on twistor space, in which case they correspond to the chart
\eq{\begin{array}{@{}c@{}}\displaystyle f_1\equiv\frac{\ab{1256}\ab{1678}\ab{2346}}{\ab{1236}\ab{1456}\ab{2678}}\,,\;\; f_2\equiv\frac{\ab{1678}\ab{2356}}{\ab{1256}\ab{3678}}\,,\;\; f_3\equiv\frac{\ab{1235}\ab{1268}}{\ab{1238}\ab{1256}}\,,\\[10pt]\displaystyle f_4\equiv\frac{\ab{1236}\ab{1567}\ab{2678}}{\ab{1267}\ab{1678}\ab{2356}}\,,\qquad f_5\equiv\frac{\ab{1256}\ab{1678}}{\ab{1268}\ab{1567}}\,.\end{array}\label{octagon_a_coordinate_chart}}
(The attentive reader will notice that, although the plabic graph \eqref{octagon_a_plabic_graph} has a smooth degeneration to \eqref{heptagon_b_plabic_graph} and the twistors \eqref{octagon_a_boundary_measurements} to \eqref{heptagon_b_boundary_measurements}, the chart \eqref{octagon_a_coordinate_chart} does not obviously degenerate to \eqref{heptagon_b_bespoke_coordinates} smoothly (although it does). But as already discussed, this limit of octagon A would not result in heptagon B---again, due to the pentabox contributions to octagon A which would diverge in this limit.)

As before, there is an obvious generalization of (\ref{octagon_a_dual_sequence}) to arbitrary loop order:
\eq{\fig{-29.67pt}{1}{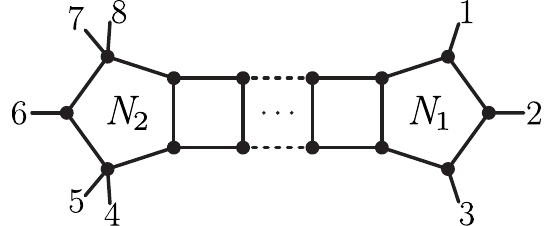}.\label{octagon_a_ladder_integral}}
We have performed these integrals through four loops, and we have included the result through three loops as ancillary files. We find that the octagon A ladders can be expressed as polynomials in $\log (f_2)$ whose coefficients are polylogarithms drawn from the basis
\begin{align}\bigg\{ G_{\vec{w}}(f_1) \bigg| w_i\!\in\!&\left\{0, \frac{\mi1}{1\pl f_2} \right\}\! ,\nonumber\\
G_{\vec{w}}(f_4) \bigg| w_i\!\in\!&\left\{0, \mi 1,\frac{\mi1}{1\pl f_1},\mi\frac{1\pl f_2}{f_2} \right\} \!,\nonumber\\
 G_{\vec{w}}(f_3) \bigg| w_i\!\in\!&\,\bigg\{0, \mi 1,\frac{\mi1}{1\pl f_4}, \mi\frac{1\pl f_1}{f_1}, \mi\frac{1\pl f_2}{1\pl f_2(1\pl f_4)}, \mi\frac{1\pl f_1(1\pl f_2)}{f_1 (1\pl f_2)},\mi\frac{1\pl f_1(1\pl f_2)}{f_1(1\pl f_2(1\pl f_4))} \bigg\},\nonumber\\
 G_{\vec{w}}(f_5) \bigg| w_i\!\in\!&\,\bigg\{0, \mi 1, \mi\frac{1\pl f_4}{f_4}, \mi\frac{1\pl f_3(1\pl f_4)}{f_3 f_4},\mi\frac{1\pl f_2(1\pl f_4)}{f_2 f_4},\mi\frac{1\pl f_4(1\pl f_1)}{f_4 (1\pl f_1)},\\
&\quad\mi\frac{(1\pl f_3(1\pl f_4))(1\pl f_2(1\pl f_4))}{f_2 f_3 f_4(1\pl f_4)},\mi\frac{1\pl f_2\pl f_3\pl f_2 f_3(1\pl f_4)}{f_2 f_3 f_4},\nonumber\\
&\quad\mi\frac{1\pl f_1(1\pl f_2\pl f_3\pl f_2 f_3(1\pl f_4))}{f_1 f_2 f_3 f_4} \bigg\}\!\bigg\} .\nonumber\end{align}
Three-loop integrals in this class take under an hour, while four-loop integrals take under a day. In summary, we find that the machinery for heptagon B goes through virtually unchanged for octagon A.

\vspace{-2pt}
\subsubsection{Octagon B: Kinematic Novelties at Eight Points}\label{subsubsec:octagonal_novelties_kinematic}%
\vspace{-2pt}

New features occur for the second integral in \eqref{two_loop_eight_point_topologies}. We can cut to the chase slightly, and immediately consider its $L$-loop ladder version:
\eq{\fig{-29.67pt}{1}{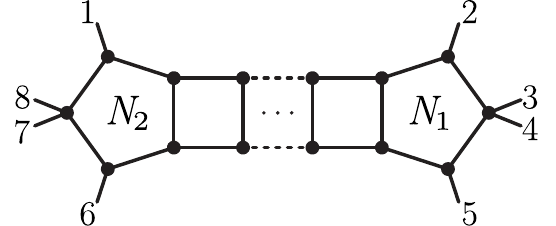},\label{octagon_b_ladder_integral}}
where the numerators are given in \eqref{octa_ladder_numerators_appendix} of \mbox{appendix \ref{appendix:derivation_of_the_octagon_ladder}}. Moreover,  \mbox{appendix \ref{appendix:derivation_of_the_octagon_ladder}} includes a detailed illustration of how the Feynman parameterization of this ladder integral can be obtained following the methods described in \mbox{ref.\ \cite{conformalIntegration}}.

This family of integrals depends on the five cross-ratios
\eq{\hspace{-85pt}u_1\equiv\u{13}{57}\,,\quad u_2\equiv\u{25}{61}\,,\quad u_3\equiv\u{36}{72}\,,\quad u_4\equiv\u{35}{62}\,,\quad u_5\equiv\u{71}{26}\,.\hspace{-50pt}\label{octagon_b_cross_ratios}}
Obviously, octagon B generalizes the family of heptagon A ladders treated in subsection \ref{subsubsec:heptagon_a_section}; in particular, it smoothly degenerates to them.
Thus, we would like to choose coordinates that smoothly degenerate to the coordinates in subsection \ref{subsubsec:heptagon_a_section}.
Concretely, we may choose to parameterize momentum twistors according to the edge chart of the following plabic graph:
\eq{\hspace{-10pt}\fig{-42.185pt}{1}{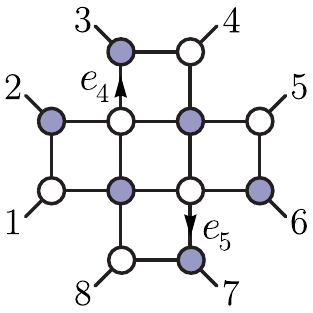}\;\;Z^{(8)}_{\text{B}}\!\equiv\!\!\raisebox{-0pt}{$\left(\raisebox{28pt}{\!}\right.$}\begin{array}{@{}c@{$\;$}c@{$\;$}c@{$\;$}c@{$\;$}c@{$\;$}c@{$\;$}c@{$\;$}c@{}}1&0&0&e_4&0&\mi\!1\phantom{\mi\!}&\mi\!1\mi e_2(1\pl e_3)&\mi\!1\mi (e_2\pl e_5)(1\pl e_3)\\[-2pt]1&1&0&0&0&0&\mi e_2e_3\phantom{\mi}&\mi e_3(e_2\pl e_5)\\[-2pt]0&0&1&1\pl e_1e_4&1&1&0&0\\[-2pt]0&0&0&0&1&1\pl e_1&e_1&e_1\end{array}\raisebox{-0pt}{$\left.\raisebox{28pt}{\!}\right).$}\vspace{0pt}\label{octagon_b_plabic_graph_and_boundary_measurements}}
Viewed as coordinates on twistor space, this corresponds to the chart
\eq{\begin{array}{@{}c@{}}\displaystyle e_1\equiv\frac{\ab{1237}\ab{1256}}{\ab{1235}\ab{1267}}\,,\;\; e_2\equiv\frac{\ab{1235}\ab{1567}\ab{2456}}{\ab{1257}\ab{1456}\ab{2356}}\,,\;\; e_3\equiv\frac{\ab{1256}\ab{4567}}{\ab{1567}\ab{2456}}\,,\\[10pt]\displaystyle e_4\equiv\frac{\ab{1267}\ab{2345}}{\ab{1237}\ab{2456}}\,,\qquad e_5\equiv\frac{\ab{1235}\ab{1256}\ab{1678}}{\ab{1257}\ab{1268}\ab{1356}}\,.\end{array}\label{octagon_b_bespoke_coordinate_chart}}
One can immediately see that the above expressions smoothly degenerate to \eqref{heptagon_a_bespoke_chart_with_boundary_measurements} and  \eqref{heptagon_a_bespoke_coordinates} for $z_7\!\parallel\!z_8$, as $z_8$ only occurs in $e_5$, which vanishes in this limit.

In this coordinate chart \eqref{octagon_b_bespoke_coordinate_chart}, the five cross-ratios (\ref{octagon_b_cross_ratios}) become
\eq{\begin{array}{@{}l@{$\;\;\;\;$}l@{}}\displaystyle u_1\!=\!\frac{e_1e_2e_3}{(1\pl(1\pl e_1)e_2(1\pl e_3))(1\pl(1\pl e_1 e_4)(e_2\pl e_5))}\,,&\displaystyle u_2\!=\!\frac{(1\pl e_1 e_4)(e_2\pl e_5)}{1\pl(1\pl e_1e_4)(e_2\pl e_5)}\,,\\[14pt]\displaystyle u_3\!=\!\frac{1}{1\pl(1\pl e_1)e_2(1\pl e_3)}\,,&\hspace{-85pt}\displaystyle u_4\!=\!\frac{e_1e_4}{1\pl e_1e_4}\,,\hspace{20pt} u_5\!=\!\frac{e_5}{e_2\pl e_5}\,.\end{array}\label{octagon_b_us_in_es}}
As always with momentum twistors, it is not hard to verify that the $6\!\times\!6$ Gramian determinant is rationalized in these coordinates:
\eq{\begin{split}\gramian{8}{\{123567\}}&=\sqrt{(1\mi u_1\mi u_2\mi u_3\pl u_2u_3(u_4\pl u_5\mi u_4u_5))^2\mi 4 u_1u_2u_3(1\mi u_4)(1\mi u_5)}\\&=\frac{e_2(e_1\mi e_3)}{(1\pl(1\pl e_1)e_2(1\pl e_3))(1\pl(1\pl e_1 e_4)(e_2\pl e_5))}\,.\end{split}\label{octagon_b_gramian_in_es}}
However, it turns out that upon direct integration (which is not algorithmically any more difficult than for heptagon A, in the sense that heptagon C is), another kinematic square root arises that is not trivially rationalized in momentum twistor coordinates.

\paragraph{The `Four-Mass' Square Root for Eight-Point Kinematics}~\\[-14pt]

The additional square root we encounter when we integrate octagon B is
\eq{\gramian{8}{\{1357\}}=\sqrt{(1\mi u_1\mi u_2u_3u_4u_5)^2\mi 4 u_1u_2u_3u_4u_5}\,,\label{octagon_b_four_mass_gramian}}
where we have introduced the notation
\eq{\gramian{n}{A}\equiv\sqrt{\det\!\big(G^{A}_{\phantom{A}A}\big)/\big(\x{a_1}{a_3}^2\x{a_2}{a_4}^2\big)}\quad\text{for } A\!=\!\{a_1,\ldots,a_4\}\,.\label{definition_of_gramian_4_determinants_in_x_space}}
It is the square root of a $4\!\times\!4$ Gramian determinant! 

One way to see that this square root is important for the integral is to observe that octagon B has support on a specific residue: we cut the propagators $(\ell_1,2)$, $(\ell_1,4)$, $(\ell_2,6)$, $(\ell_2,8)$ and $(\ell_1,\ell_2)$; this yields a Jacobian that we cut, which yields a second Jacobian that we cut; which yields one final Jacobian that we also cut to uncover the square root for a $4\!\times\!4$ Gramian determinant. 
(When computing the MHV two-loop amplitude, this square root will cancel in the sum of the present integral and its image under a cyclic shift of the external legs by two. However, this root is certainly relevant to amplitudes beyond MHV.)

The effect of taking the residue described above is that, on the residue, $\ell_1\!=\!\ell_2$ and both are a solution to the `quad-cut' equations associated with the four propagators mentioned. This is extremely reminiscent of what happens at one loop. Indeed, recalling that $u_1\!\equiv\!\u{13}{57}$, and observing that 
\eq{u_2 u_3u_4 u_5=\u{35}{71}\,,\label{four_mass_v_in_us}}
it is clear that (\ref{octagon_b_four_mass_gramian}) is just the square root of the famous `four-mass' box integral \cite{Hodges:boxInt,Hodges:2010kq,Mason:2010pg},
\eq{\fig{-29.67pt}{1}{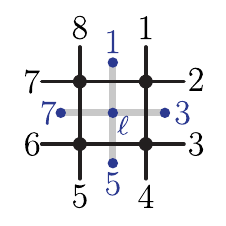}\propto\int\limits_0^\infty\!\!d^2\!\vec{\alpha}\frac{1}{(1\pl\alpha_1 u\pl\alpha_2 v)(\alpha_1\pl\alpha_2\pl\alpha_1\alpha_2)}\,,\label{four_mass_box_parametric_integral}}
where $u\!\equiv\!\u{13}{57}$ and $v\!\equiv\!\u{35}{71}$. In this case, the way to rationalize this root is also quite well known: one merely trades $(u,v)$ for the `light-cone' coordinates $z,\bar{z}$ according to:
\eq{ u=z\bar{z}\,,\qquad v=(1-z)(1-\bar{z})\,,}
which renders $\gramian{8}{\{1357\}}=\pm(z\mi\bar{z})$.

We could perform an analogous ad-hoc change of variables on top of $\{u_i\}\!\to\!\{e_i\}$ in order to rationalize also $\gramian{8}{\{1357\}}$. More generally, however, it is not clear how to discover parameterizations for kinematics which rationalize {\it all} the potential algebraic roots encountered through integration. (As we will see in the conclusions, this is a much more difficult task than merely rationalizing the $6\!\times\!6$ and $4\!\times\!4$ Gramian determinants.)

Even without rationalizing this final four-mass square root, however, we were able to integrate the two-loop octagon B using {\tt HyperInt} in less than an hour. As the appearance of these kinds of kinematic square roots poses an interesting challenge to symbology, for example, we expect it to be valuable for other researchers. We include the two-loop expression as an ancillary file to this work's submission.

\vspace{-2pt}
\subsubsection{Octagon K: Algorithmic Novelties at Eight Points}\label{subsubsec:octagonal_novelties_algorithmic}
\vspace{-2pt}

Finally, let us address the last two-loop octagon (`K') integral in \eqref{two_loop_eight_point_topologies}:
\eq{\fwbox{0pt}{\fig{-29.67pt}{1}{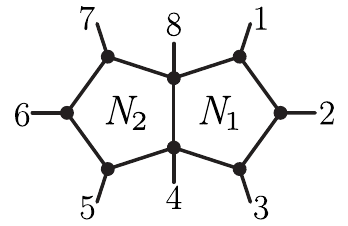}\bigger{\Leftrightarrow}\fig{-29.67pt}{1}{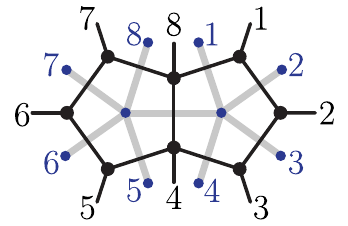}\bigger{\Leftrightarrow}\fig{-29.67pt}{1}{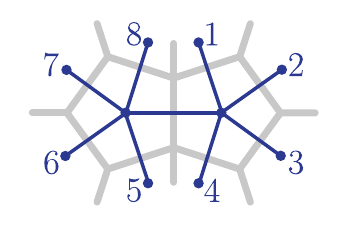}\fwboxL{0pt}{.}}\label{octagon_k_dual_sequence}}
In contrast to the previous two examples, it depends on all nine independent cross-ratios that can be formed for eight massless particles.
In parameter count, it is thus halfway between the previous two octagons considered and the most general integral occurring for two-loop MHV amplitudes, which depends on 13 parameters. (In algorithmic complexity, it seems considerably closer to the general case than to octagon B.)

Starting with a five-fold Feynman-parametric representation of this integral in terms of cross-ratios (or a twistor-parametrization thereof), a na\"{i}ve attempt at direct integration shows the problem to be substantively worse than the one described in subsection \ref{subsubsec:heptagon_c_section}. After only two integrations, one obtains an irreducible quadratic polynomial in all three remaining integration parameters. Partial fractioning any one would result in square roots involving both of the final integration parameters. This is yet another example of the general obstruction discussed in \mbox{section \ref{subsec:partial_fractioning_roots}}; in the notation of \eqref{eq: partial fractioning}, the square root $\sqrt{f^2\mi g}$ would now depend on two final integration variables. A systematic approach to rationalizing these Feynman-parameter-dependent square roots is clearly an important problem, which we must leave to future research.

\newpage\vspace{-6pt}
\section{Conclusions and Future Directions}\label{sec:conclusions}%
\vspace{-10pt}

In this paper, we have shown how loop integrals can be non-redundantly parametrized in momentum-twistor space, and that such parameterizations rationalize many of the square roots that naturally occur in cross-ratio space. This makes a large class of integrals directly amenable to Feynman-parametric integration via partial fractioning. In particular, we have illustrated how this works in a number of examples, including integrals involving up to eight particles and up to four loops. 

While momentum twistors automatically rationalize square roots of \mbox{$6\!\times\!6$} Gramian determinants, they do not rationalize square roots of \mbox{$4\!\times\!4$} Gramian determinants.
The question as to whether such square roots can be systematically rationalized is worth further investigation. Of course, Gramian square roots are not the only algebraic roots that should be relevant to loop integrals. Indeed, it is well known that roots of arbitrary high order can arise in sufficiently high-loop Feynman integrals \cite{ArkaniHamed:book}. The first example involving a cube root, for instance, arises at three loops and at least eleven particles:
\eq{\fig{-42.185pt}{1}{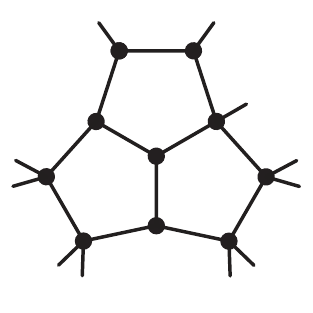}\,.\label{simplest_integral_requiring_cube_roots}}
This integral can be seen to have (maximal codimension) residues that depend on cube roots of the kinematic cross-ratios. These are clearly in a different class than those related to the Gramian matrix discussed in this work. Understanding how to rationalize these higher roots would be an interesting direction for future research. 

Beyond purely kinematic roots, the appearance of algebraic quantities involving integration parameters presents a barrier to algorithmic integration. We have identified examples of this type of obstruction at both seven and eight points. It is hoped that further study of these examples will (alongside similar examples~\cite{Panzer:2014gra}) lead to a more systematic understanding of how these barriers can be resolved more generally. In cases where Feynman-parameter-dependent square roots can no longer be avoided, elliptic integrals \mbox{\cite{Bourjaily:2017bsb,Broadhurst:1993mw,Bloch:2013tra,Broadhurst:2016myo,Bogner:2017vim}} and integrals of even higher complexity occur \mbox{\cite{Bloch:2014qca,Bloch:2016izu,Bourjaily:2018ycu}}.
These will require a new perspective on direct integration, rooted in a deeper understanding of the functions required.

Finally, let us mention that our heptagon integral results already allow us to address some concrete physical questions. In particular, they are the most complicated ingredients of planar two-loop seven-point amplitudes, both for MHV and NMHV. While the remainder function and the symbol of the ratio function are known at seven points~\cite{Golden:2014xqf,CaronHuot:2011kk}, calculating the amplitude directly from integrals at this multiplicity will prove instructive for pursuing this approach at higher points \cite{us_in_the_future}.

\vspace{\fill}\vspace{-4pt}
\section*{Acknowledgements}%
\vspace{-4pt}
We are grateful to John Golden, Johannes Henn, Enrico Herrmann, Marcus Spradlin and  Erik Panzer for illuminating discussions; and to Johannes Henn, Enrico Herrmann, and Julio Parra-Martinez for discussing intermediate results obtained using other methods for some of the integrals discussed here. We are especially grateful to Erik Panzer for concrete help with the use of {\tt HyperInt}. We thank Marcus Spradlin for comments on the manuscript.
We thank the Institute for Advanced Study in Princeton for kind hospitality during the initial phase of this project.
This work was supported in part by the Danish Independent Research Fund under grant number DFF-4002-00037 (MW), the Danish National Research Foundation (DNRF91), a grant from the Villum Fonden, and a Starting Grant \mbox{(No.\ 757978)} from the European Research Council (JLB,AJM,MvH,MW).

\newpage\appendix\vspace{-6pt}
\section[\mbox{Feynman Parameterization of an All-Loop Octagon Ladder}]{\mbox{\hspace{0pt}Parametrization of an All-Loop Octagon Ladder Integral}}\label{appendix:derivation_of_the_octagon_ladder}%
\vspace{-6pt}

For each of the examples discussed in this work, we were able to exploit an explicitly dual-conformal Feynman-parametric representation of the integral. By this, we mean a representation free of reference to (non-conformal) Mandelstam invariants $\x{a}{b}$, but only dual-conformal cross-ratios. (Moreover, the attentive reader may have noticed that all of our examples' denominators depended exclusively on `parity-even' cross-ratios---those rationally related to the $\u{ab}{cd}$'s described in \mbox{section \ref{subsec:dual_momenta_and_conformal_invariance}}.) 

The construction of such parametric representations is not the subject of our present work. Interested readers should consult \mbox{ref.\ \cite{conformalIntegration}}, where the general strategy is described in considerable detail. However, we expect that an additional concrete example (or two) would prove useful to some readers. Therefore, in this appendix, we derive the manifestly DCI Feynman-parametric representation of one of the all-loop octagon ladder integrals relevant to MHV amplitudes. We chose this example in part because it smoothly degenerates to the case of seven (or even six) particles. The Feynman-parametric representations of other integrals described in this work are obtainable through essentially the same methods described here; we leave those parameterizations as exercises for the interested readers. 

We are interested in obtaining a (manifestly dual-conformally invariant) Feynman parameterization of the following class of integrals:
\eq{I_{8,\text{B}}^{(L)}\equiv\fig{-29.67pt}{1}{octagon_b_ladder_integral}\,.\label{octagon_b_ladder_for_appendix}}
Expressed in loop-momentum space (in dual coordinates), this corresponds to the integral
\eq{\hspace{-20pt}I_{8,\text{B}}^{(L)}\equiv\!\!\int\!\!d^{4L}\vec{\ell}\frac{\x{\ell_1}{N_1}\x{2}{6}^{L-1}\x{\ell_L}{N_2}}{\x{\ell_1}{3}\x{\ell_1}{5}\big(\prod_{i=1}^L\x{\ell_i}{2}\x{\ell_i}{6}\big)\big(\x{\ell_1}{\ell_2}\cdots\x{\ell_{L-1}}{\ell_L}\big)\x{\ell_L}{7}\x{\ell_L}{1}}\,.\hspace{-20pt}\label{octa_ladder_loop_space_definition}}
For the sake of concreteness, we consider only the tensor numerators $N_i$ relevant to the MHV amplitudes (\ref{mhv_double_pentagon_loop_space_integrand}). (The analysis for the case of numerators with `mixed' parity---analogous to $\widetilde{\Omega}^{(L)}$ for six particles---is essentially identical to what follows.) Thus, we consider the case when the numerators $N_i$, expressed in twistor space, correspond to the lines denoted
\eq{N_1\!\Leftrightarrow\! (123)\tcap(456)\,,\qquad N_2\!\Leftrightarrow\!(567)\tcap(812)\,.\label{octa_ladder_numerators_appendix}}

Let us decompose the integrand of (\ref{octa_ladder_loop_space_definition}), denoted $\mathcal{I}$, according to the factors' dependence on $\ell_1$: \mbox{$\mathcal{I}\!\equiv\!\mathcal{I}_1(\ell_1)\!\times\!\mathcal{I}'(\ell_2,\ldots,\ell_L)$} where $\mathcal{I}_1$ denotes all factors involving $\ell_1$. (Notice that $\mathcal{I}_1(\ell_1)$ does also depend on $\ell_2$ because of the internal propagator.) 

This $\ell_1$-dependent part of (\ref{octa_ladder_loop_space_definition}), $\mathcal{I}_1(\ell_1)$, is a standard one-loop pentagon integral. Following the methods described in ref.\ \cite{conformalIntegration}, we introduce Feynman parameters according to
\eq{Y_1\equiv \underbrace{{\color{black}\beta_1}(3)\pl{\color{black}\beta_2}(5)\pl\alpha_1^1(2)\pl\alpha_2^1(6)}_{(Q_1)}\pl{\color{hred}\gamma_1}(\ell_2)\,,\label{first_y_definition}}
so that
\eq{\int\!\!d^4\ell_1\,\mathcal{I}_1(\ell_1)\propto\int\limits_0^{\infty}\!\!d^2\!\vec{\alpha}\proj{d\beta}d{\color{hred}\gamma_1}\,\frac{\x{Y_1}{N_1}}{\x{Y_1}{Y_1}^3}\,.\label{initial_first_loop_param_integral}}
The choices we have made in denoting the Feynman parameters in (\ref{first_y_definition}) are admittedly awkward---and so is our choice to consider the projective redundancy among the complete set of Feynman parameters to be restricted entirely to the $\beta_i$'s in (\ref{initial_first_loop_param_integral}). As the reader may have already guessed, these choices  will ensure a nice recursive notational structure that we will discover momentarily. 

Notice that the Feynman parameter integral in (\ref{initial_first_loop_param_integral}) simplifies considerably because $\x{Y_1}{Y_1}$ is linear in ${\color{hred}\gamma_1}$, and because 
\eq{\x{Y_1}{N_1}={\color{hred}\gamma_1}\x{\ell_2}{N_1}\,.}
Thus, the ${\color{hred}\gamma_1}$ integration is simple. Up to non-kinematic pre-factors that play no role in our analysis, we immediately see that
\eq{\int\!\!d^4\ell_1\,\mathcal{I}_1(\ell_1)\propto\int\limits_0^{\infty}\!\!d^2\!\vec{\alpha}\proj{d\beta}\,\frac{\x{\ell_2}{N_1}}{\x{Q_1}{Q_1}\x{\ell_2}{Q_1}^2}\,.\label{first_loop_param_integral}}

After combining (\ref{first_loop_param_integral}) with the rest of the integrand, $\mathcal{I}'(\ell_2,\ldots,\ell_L)$, we see that the $\ell_2$ part of the integral is virtually identical in form to the first. Indeed, the primary difference is merely that the $\ell_2$ integral involves only four propagators, with one propagator squared. Introducing Feynman parameters according to\footnote{Notice that we have fixed the projective redundancy here to ensure the coefficient of $(Q_1)$ in (\ref{y2_of_octa_ladder}) to be 1. (Careful readers will notice that this ensures we do not need any power of this would-be Feynman parameter in the numerator of the resulting Feynman-parametric integral.)}
\eq{Y_2\equiv\underbrace{(Q_1)\pl\alpha_1^2(2)\pl\alpha_2^2(6)}_{(Q_2)}\pl{\color{hred}\gamma_2}(\ell_3)\,,\label{y2_of_octa_ladder}}
it is easy to see that upon integrating ${\color{hred}\gamma_2}$ as above
\eq{\int\!\!d^4\ell_2\frac{\x{\ell_2}{N_1}}{\x{\ell_2}{Q_1}^2\x{\ell_2}{2}\x{\ell_2}{6}\x{\ell_2}{\ell_3}}\propto\int\limits_0^{\infty}\!\!d\alpha_1^2d\alpha_2^2\frac{\x{\ell_3}{N_1}}{\x{Q_2}{Q_2}\x{\ell_3}{Q_2}^2}\,.}

Continuing in this manner through $\ell_{L-1}$, we see that the original integral (\ref{octa_ladder_loop_space_definition}) is proportional to 
\eq{\int\limits_0^{\infty}\!\!\frac{d^{2(L-1)}\!\vec{\alpha}\proj{d\beta}\hspace{10pt}\x{2}{6}^{L-1}}{\x{Q_1}{Q_1}\cdots\x{Q_{L-1}}{Q_{L-1}}}\int\!\!d^4\!\ell_L\,\frac{\x{\ell_L}{N_1}\x{\ell_L}{N_2}}{\x{\ell_L}{Q_{L-1}}^2\x{\ell_L}{1}\x{\ell_L}{2}\x{\ell_L}{6}\x{\ell_L}{7}}\,.\label{penultimate_octa_ladder}}
Here, each $(Q_k)$ appearing above is defined recursively in the obvious way:
\eq{\begin{split}(Q_0)&\equiv{\color{black}\beta_1}(3)\pl{\color{black}\beta_2}(5)\,,\\
(Q_k)&\equiv (Q_{k-1})\pl\alpha_1^k(2)\pl\alpha_2^k(6)\,.\end{split}}

The final loop integration to perform in (\ref{penultimate_octa_ladder}) is essentially a one-loop `hexagon' integral (where one propagator is repeated). Because there is a massless leg involved, we can always write this as a two-fold Feynman parametric integral. To see this, let us introduce Feynman parameters according to\footnote{We apologize to the attentive for the unanticipated denotation of `$\alpha_2^L$' as the coefficient of $(1)$.}
\eq{Y_L\equiv\underbrace{(Q_{L-1})\pl\alpha_1^L(2)\pl\alpha_2^{L}(1)}_{(R)}\pl{\color{hred}\eta_1}(6)\pl{\color{hred}\eta_2}(7)\,.\label{final_y_of_octa}}
The motivation for this final representation follows from the fact that (the masslessness of $p_7$ ensures that) there is no term proportional to ${\color{hred}\eta_1\,\eta_2}$ in $\x{Y_L}{Y_L}$, and thus both parameters can be integrated out algebraically as total derivatives. The motivation behind this notation follows from the observation that
\eq{\begin{split}&\int\!\!d^4\!\ell_L\,\frac{\x{\ell_L}{N_1}\x{\ell_L}{N_2}}{\x{\ell_L}{Q_{L-1}}^2\x{\ell_L}{1}\x{\ell_L}{2}\x{\ell_L}{6}\x{\ell_L}{7}}\\
\propto&\int\limits_0^\infty\!\!d\alpha_1^Ld\alpha_2^Ld^2{\color{hred}\vec{\eta}}\,\left(3\frac{\x{Y_L}{N_1}\x{Y_L}{N_2}}{\x{Y_L}{Y_L}^4}\mi\frac{\x{N_1}{N_2}}{\x{Y_L}{Y_L}^3}\right),\end{split}\label{penultimate_octa_ladder_2}}
and that the separate linearity of $\x{Y_L}{Y_L}$ in each ${\color{hred}\eta_{1,2}}$  (with no mixed term) ensures a fairly standard form of the answer after the ${\color{hred}\eta_i}$-integrals. The first step to see this clearly is to notice that, as a consequence of the way in which the chiral numerators $N_i$ have been defined in (\ref{octa_ladder_numerators_appendix}), $\x{Y_L}{N_1}\x{Y_L}{N_2}$ simplifies considerably:
\eq{\begin{split}\x{Y_L}{N_1}\x{Y_L}{N_2}&=\big(\x{R}{N_1}\pl{\color{hred}\eta_2}\x{7}{N_1}\big)\x{Q_0}{N_2}\\
&=\alpha_2^L\x{1}{N_1}\x{Q_0}{N_2}\pl{\color{hred}\eta_2}\x{7}{N_1}\x{Q_0}{N_2}\,.\end{split}\label{penult_numerator_pieces}}
This allows us to recognize the ${\color{hred}\eta_i}$ integrals in (\ref{penultimate_octa_ladder_2}) as being of the form
\eq{\int\limits_0^\infty\!\!d^2{\color{hred}\vec{\eta}}\left(\!3\frac{n_1\pl n_2{\color{hred}\eta_2}}{(g_1\pl g_2{\color{hred}\eta_2}\pl g_3{\color{hred}\eta_1})^4}-\frac{n_0}{(g_1\pl g_2{\color{hred}\eta_2}\pl g_3{\color{hred}\eta_1})^3}\right)\propto\frac{1}{g_1g_2g_3}\left(\frac{n_1}{g_1}\pl\frac{n_2}{g_2}\mi n_0\right).\label{schematic_final_integration}}
Recognizing the terms in (\ref{schematic_final_integration}) from the numerator (\ref{penult_numerator_pieces}) and denominator ingredients in (\ref{final_y_of_octa}) and including the parts from previous integrations (and including numerical constants of proportionality we have mostly ignored in our analysis so far), we see that we may conclude that
\eq{\hspace{-20pt}I_{8,\text{B}}^{(L)}\equiv\!\!\int\limits_0^\infty\!\!d^{2L}\!\vec{\alpha}\proj{d\beta}\frac{\x{2}{6}^{L-1}}{(f_1\cdots f_{L-1})\,g_1\,g_2\,g_3}\left(\frac{n_1}{g_1}\pl\frac{n_2}{g_2}\mi n_0\right),\hspace{-20pt}\label{octa_ladder_penultimate_form}}
where we have defined the ingredient functions
\eq{\begin{split}f_k\equiv\frac{1}{2}\x{Q_k}{Q_k}\,,\hspace{19.5pt} g_1\equiv\frac{1}{2}\x{R}{R}\,,\hspace{19.5pt} g_2\equiv\x{R}{7}\,,\hspace{19.5pt} g_3\equiv\x{R}{6}\,,\\
n_0\equiv\x{N_1}{N_2}\,,\quad n_1\equiv \alpha^{L}_2\x{R}{N_1}\x{Q_0}{N_2}\,,\quad n_2\equiv \x{7}{N_1}\x{Q_0}{N_2}\,.\end{split}}

The final step to achieve the representation used in \mbox{section \ref{subsec:octagons_and_beyond}} is to rescale the Feynman parameters in order to render (\ref{octa_ladder_penultimate_form}) in a form which is manifestly dual-conformally invariant. This will be achieved by 
\eq{\begin{array}{@{}c@{}}\displaystyle\alpha^k_1\!\mapsto\!\alpha^k_1/\x{2}{6}\,,\quad \alpha^{k<L}_2\!\!\mapsto\!\alpha^k_2\frac{\x{1}{3}\x{2}{7}}{\x{1}{6}\x{2}{6}\x{3}{7}}\,,\quad \alpha^L_2\!\mapsto\!\alpha^L_2/\x{1}{6}\,,\\
\displaystyle\beta_1\!\mapsto\!\beta_1\frac{\x{2}{7}}{\x{2}{6}\x{3}{7}}\,,\hspace{96pt}\beta_2\!\mapsto\!\beta_2\frac{\x{1}{3}\x{2}{7}}{\x{1}{5}\x{2}{6}\x{3}{7}}\,.\end{array}\label{append_ocatagon_a_parametric_rescalings}}
(Recall that $\alpha^L_2$, a Feynman parameter associated with the dual point $(1)$, is actually quite different from $\alpha^{k<L}_2$, which are associated with the dual point $(6)$.) 

Under this rescaling, the factors in the denominator of (\ref{schematic_final_integration}) all become uniform in conformal weights. Because the denominators are uniform in weight, they are conformal up to a factor which can be absorbed into the numerator. After taking all these factors into account together with the Jacobians required by the rescalings (\ref{append_ocatagon_a_parametric_rescalings}) (and allowing for a slight abuse of notation), we obtain the following representation of the integral:
\eq{\hspace{-20pt}I_{8,\text{B}}^{(L)}\equiv\!\!\int\limits_0^\infty\!\!d^{2L}\!\vec{\alpha}\proj{d\beta}\frac{u_1}{(f_1\cdots f_{L-1})\,g_1\,g_2\,g_3}\left(\frac{\alpha^L_2(\beta_1n^1_1\pl \beta_2n^1_2)}{g_1}\pl\frac{\beta_1n^2_1\pl\beta_2n^2_2}{g_2}\mi1\right),\hspace{-20pt}\vspace{-5pt}\label{octa_ladder_ultimate_form_0}}
where the denominators are given by
\vspace{-14pt}\eq{\begin{split}f_k&\!\equiv\big(\alpha_1^1\pl\!\ldots\pl\alpha_1^k\big)\beta_2 u_2\,\,\pl\,\big(\alpha_2^1\pl\!\ldots\pl\alpha_2^k\big)\beta_1 u_3\,\,\pl\,\beta_1\beta_2 u_2u_3u_4\,\,\,\pl\sum_{i,j=1}^k\!\alpha_1^i\alpha_2^j\,,\\[-14pt]
g_1&\!\equiv f_{L-1}\pl\big(\alpha_2^1\pl\!\ldots\pl\alpha_2^{L-1}\big)\big(\alpha_1^L\pl\alpha_2^L\big)\pl\alpha_1^L\beta_2\,u_2\pl\alpha_2^L\big(\beta_1\pl\beta_2\big)\,,\\
g_2&\!\equiv\big(\alpha^1_1\pl\!\ldots\pl\alpha_1^L\big)\pl\alpha_2^Lu_5\pl\beta_1\pl\beta_2 u_1\,,\qquad g_3\!\equiv\big(\alpha^1_1\pl\!\ldots\pl\alpha_1^L\big)\pl\alpha_2^L\pl\beta_1u_3\,,\\[-6pt]\end{split}\label{octagon_b_ladder_denominators_defined_penult}}
in terms of the cross-ratios (defined as in (\ref{octagon_b_cross_ratios})),
\vspace{-2pt}\eq{\hspace{-85pt}u_1\equiv\u{13}{57}\,,\quad u_2\equiv\u{25}{61}\,,\quad u_3\equiv\u{36}{72}\,,\quad u_4\equiv\u{35}{62}\,,\quad u_5\equiv\u{26}{71}\,;\hspace{-50pt}\label{octagon_b_cross_ratios_defined_appendix}}
and the numerator coefficients $n_i^j$ are given in terms of twistor four-brackets:
\eq{\fwbox{0pt}{\hspace{-30pt}n_{1}^1\!\!\equiv\!\!\frac{\ab{1\,456}\ab{2\,567}}{\ab{12\,56}\ab{45\,67}}, n_{2}^1\!\!\equiv\!\!\frac{\ab{1\,456}\ab{812\,5}}{\ab{81\,45}\ab{12\,56}}, n^2_1\!\!\equiv\!\!\frac{\ab{123\,6}\ab{2\,567}}{\ab{12\,56}\ab{23\,67}}, n^2_2\!\!\equiv\!u_1\!\frac{\ab{123\,6}\ab{812\,5}}{\ab{81\,23}\ab{12\,56}}\!.}\label{numerator_factors_as_four_brackets}}

At this stage, it would be easy to evaluate the numerators (\ref{numerator_factors_as_four_brackets}) in a coordinate chart such as that described in \mbox{section \ref{subsec:octagons_and_beyond}} (see (\ref{octagon_b_plabic_graph_and_boundary_measurements})). Moreover, evaluating the cross-ratios (\ref{octagon_b_cross_ratios_defined_appendix}) in such a chart would allow this evaluation to be inverted, resulting in an explicit form of the numerators expressed in terms of the cross-ratios $u_i$. {\it However}, as emphasized throughout this work, this would generally involve algebraic roots---specifically, the square root associated with the (only relevant) $6\!\times\!6$ Gramian determinant (see (\ref{octagon_b_gramian_in_es})). As such, it would seem that the best representation of the numerators $n_i^j$ would be that given in terms of momentum-twistor cross-ratios given above (\ref{numerator_factors_as_four_brackets}). 

However, it turns out that there is (at least) one relation among the pieces appearing in the representation (\ref{octa_ladder_ultimate_form_0}). Specifically, there is an identity 
\eq{\hspace{-20pt}\!\!\int\limits_0^\infty\!\!d^{2L}\!\vec{\alpha}\proj{d\beta}\frac{1}{(f_1\cdots f_{L-1})\,g_1\,g_2\,g_3}\left(\frac{\alpha^L_2\beta_1}{g_1}\mi\frac{u_1\beta_2}{g_2}\right)=0\,,\hspace{-20pt}\label{octa_ladder_integral_identity}}
that allows us to eliminate either $n_1^1$ or $n_2^2$. We chose the former option, changing
\eq{n_1^1\!\mapsto0\,,\qquad n_2^2\!\mapsto n_2^2\pl u_1n_1^1=u_1\left(\!\frac{\ab{123\,6}\ab{812\,5}}{\ab{81\,23}\ab{12\,56}}+\frac{\ab{1\,456}\ab{2\,567}}{\ab{12\,56}\ab{45\,67}}\!\right).}
The factor in parentheses turns out to be proportional to $1/u_1$, and independent of the square-root (\ref{octagon_b_gramian_in_es})---as was in fact the case for $n_2^1$ and $n_1^2$ from the outset. 

Thus, after exploiting the integral-level identity (\ref{octa_ladder_integral_identity}), re-expressing all coefficients directly in terms of the cross-ratios (\ref{octagon_b_cross_ratios_defined_appendix}), and eliminating the projective redundancy among the $\beta_i$ and relabeling them according to \mbox{$\{\beta_1,\beta_2\}\!\mapsto\!\{\beta,1\}$}, we find the following representation of the all-loop octagon integral (\ref{octagon_b_ladder_for_appendix}):
\eq{\hspace{-20pt}I_{8,\text{B}}^{(L)}\equiv\!\!\int\limits_0^\infty\!\!\frac{\fwboxR{30pt}{d^{2L}\!\vec{\alpha}\,d\beta\hspace{16pt}}u_1}{(f_1\cdots f_{L-1})\,g_1\,g_2\,g_3}\left(\frac{\alpha^L_2(1\mi u_2)}{g_1}\pl\frac{(\beta\pl1)(1\mi u_3)\pl u_1\mi u_2\pl u_2u_3(u_4\pl u_5\mi u_4u_5)}{g_2}\mi1\right)\!,\hspace{-20pt}\label{octa_ladder_ultimate_form}\vspace{-15pt}}
where the denominators,
\vspace{-14pt}\eq{\begin{split}f_k&\!\equiv\big(\alpha_1^1\pl\!\ldots\pl\alpha_1^k\big) u_2\,\,\pl\,\big(\alpha_2^1\pl\!\ldots\pl\alpha_2^k\big)\beta u_3\,\,\pl\,\beta u_2u_3u_4\,\,\,\pl\sum_{i,j=1}^k\!\alpha_1^i\alpha_2^j\,,\\[-14pt]
g_1&\!\equiv f_{L-1}\pl\big(\alpha_2^1\pl\!\ldots\pl\alpha_2^{L-1}\big)\big(\alpha_1^L\pl\alpha_2^L\big)\pl\alpha_1^Lu_2\pl\alpha_2^L\big(1\pl\beta\big),\\
g_2&\!\equiv u_1\pl\big(\alpha^1_1\pl\!\ldots\pl\alpha_1^L\big)\pl\alpha_2^Lu_5\pl\beta,\hspace{10pt} g_3\!\equiv\big(\alpha^1_1\pl\!\ldots\pl\alpha_1^L\big)\pl\alpha_2^L\pl\beta u_3\,,\\[-6pt]\end{split}\label{octagon_b_ladder_denominators_defined_final}}
are the same as defined above in (\ref{octagon_b_ladder_denominators_defined_penult}) after the replacement \mbox{$\{\beta_1,\beta_2\}\!\mapsto\!\{\beta,1\}$}. 

This representation smoothly degenerates to seven or six particle cases upon setting one or both of $\{u_4,u_5\}$ to zero, respectively.

\newpage\vspace{-6pt}
\section[\mbox{Explicit Numerators for \texorpdfstring{N$^{k>0}$MHV}{N(k>0)MHV} Integral Examples}]{\mbox{\hspace{0pt}Explicit Numerators for \texorpdfstring{N$^{k>0}$MHV}{N(k>0)MHV} Integral Examples}}\label{appendix:nonmhv_numerator_details}%
\vspace{-6pt}

As discussed in \mbox{section \ref{sec:illustrations_of_integration}}, the tensor numerators of 
\eq{\fig{-29.67pt}{1}{heptagon_b_integral}\quad\text{or}\quad\fig{-29.67pt}{1}{octagon_a_integral}\label{two_loop_exempli_for_appendix}}
are not uniquely defined by the criterion that these integrals be infrared finite. This is semi-obvious: both integrals admit some number of infrared-finite pentabox sub-topologies (one for the heptagon and two for the octagon in (\ref{two_loop_exempli_for_appendix})). Thus, arbitrary rational combinations of these finite sub-topologies can be added or subtracted without making these integrals infrared divergent. Correspondingly, for the sake of giving concrete expressions for these integrals (which we do give in the ancillary files for this work's submission to the {\tt arXiv}), we must be clear how these numerators are defined.

It is not possible to choose numerators for the integrals (\ref{two_loop_exempli_for_appendix}) so that they have vanishing residues on all their pentabox sub-topologies. However, one can add or subtract pentabox integrals to modify the cuts supported. One natural choice would be to choose the numerators $N_i$ so that the integrals were {\it chiral} on all pentabox cuts. This is in fact what happens for the numerators defined for the two-loop integrals relevant to MHV amplitudes (\ref{mhv_double_pentagon_loop_space_integrand}). Making a similar choice for the integrals (\ref{two_loop_exempli_for_appendix}) would have the effect of rendering their degenerations smooth. However, this is not the choice made by the authors of \mbox{ref.\ \cite{Bourjaily:2015jna}} for the purpose of representing amplitudes. Whether or not the choice made there is optimal can be debated (see the discussion in \mbox{ref.\ \cite{Bourjaily:2017wjl}} for example); but we follow their choice here for the sake of concreteness and familiarity.

Thus, let us describe the form of the tensor numerators adopted in \mbox{ref.\ \cite{Bourjaily:2015jna}}. As the pentabox contact-terms which form part of the definition of the numerator are not chiral, it turns out that the different numerators are not related simply by parity. (That is, they are related by parity---but only up to conventional corrections that are parity-even.) Thus, instead of giving two possible forms of $N_1$ and two possible forms of $N_2$ separately, we list four pairs $\{N_1,N_2\}$. Moreover, because not all the finite pentaboxes of the octagon (\ref{two_loop_exempli_for_appendix}) are finite in the degeneration to the heptagon in (\ref{two_loop_exempli_for_appendix}), these sets of chiral numerators are not smoothly related to one another. As such, we must simply give four pairs of numerators separately in the two cases. 

To be clear, we consider the two-loop integrals appearing in (\ref{two_loop_exempli_for_appendix}) to be defined in the same way as in the {\sc Mathematica} package associated with \mbox{ref.\ \cite{Bourjaily:2015jna}}, according to ($2\times$) the function {\tt localLoopIntegrand\hspace{-1pt}[i,j]\hspace{-2pt}[]}---where the index pairs take values \mbox{$\{i,j\}\!=\!\{1,1\},\{1,2\},\{2,1\}, \{2,2\}$}, defining the four possible numerators $1,\ldots,4$.

\newpage
\paragraph{Non-MHV Heptagon Ladder Numerator Details}~\\[-14pt]

We first consider the heptagon ladder integrals,
\eq{\fig{-29.67pt}{1}{heptagon_b_ladder_integral},\label{heptagon_b_ladder_appendix}}
which are defined in dual-momentum space as
\eq{\hspace{-20pt}I_{7,\text{B}}^{i,(L)}\equiv\!\!\int\!\!d^{4L}\vec{\ell}\frac{\x{\ell_1}{N_1^i}\x{1}{4}^{L-2}\x{\ell_L}{N_2^i}}{\x{\ell_1}{2}\x{\ell_1}{3}\big(\prod_{i=1}^L\x{\ell_i}{1}\x{\ell_i}{4}\big)\big(\x{\ell_1}{\ell_2}\cdots\x{\ell_{L-1}}{\ell_L}\big)\x{\ell_L}{6}\x{\ell_L}{7}}\,.\hspace{-20pt}\label{heptagon_b_ladder_loop_space_definition_appendix}}
(Notice that the factor of $\x{1}{4}^{L-2}$ in the numerator of (\ref{heptagon_b_ladder_loop_space_definition_appendix}) allows us to define the pieces $\{N_1^i,N_2^i\}$ independently of loop order.)

The four possible tensor numerator pairs defined in \mbox{ref.\ \cite{Bourjaily:2015jna}} correspond to
\vspace{-34pt}\eq{\hspace{-67pt}\begin{array}{@{}l@{}}
~\\[12pt]N_1^{1}\!\!\equiv\!\!(\hspace{-1pt}712\hspace{-1pt})\tncap(\hspace{-1pt}234\hspace{-1pt})\,,\; 
N_2^{1}\!\!\equiv\!\!\Big[\!\Big(\!(56)\tncap\big(\!(6{\color{hred}71})\tncap(13){\color{hblue}34}\big)7\!\Big)\mi\hspace{1pt}\!\Big(\!(56)\tncap\big(\!(6{\color{hblue}34})\tncap(13){\color{hred}71}\big)7\!\Big)\!\Big],\\[6pt]
N_1^{2}\!\!\equiv\!\!(\hspace{-1pt}712\hspace{-1pt})\tncap(\hspace{-1pt}234\hspace{-1pt})\,,\; 
N_2^{2}\!\!\equiv\!\!\Big[\!\Big(\!6(71)\tncap\big(\!(13)\tncap({\color{hred}34}7){\color{hblue}56}\big)\!\Big)\mi\hspace{1pt}\!\Big(\!6(71)\tncap\big(\!(13)\tncap({\color{hblue}56}7){\color{hred}34}\big)\!\Big)\!\Big],\\[6pt]
N_1^{3}\!\!\equiv\!(\hspace{-1pt}13\hspace{-1pt})\,,\; 
N_2^{3}\!\!\equiv\!\!\Big[\!\Big(\hspace{-1pt}\!\big(\!\big(\hspace{-1pt}2(\hspace{-1pt}71\hspace{-1pt})\tncap(234)\!\big)\tncap(\hspace{-1pt}6{\color{hred}71}\hspace{-1pt}){\color{hblue}34}\!\big)\tncap(\hspace{-1pt}56\hspace{-1pt})7\!\Big)\mi\hspace{1pt}\!\Big(\hspace{-1pt}\!\big(\!\big(\hspace{-1pt}2(\hspace{-1pt}71\hspace{-1pt})\tncap(234)\!\big)\tncap(\hspace{-1pt}6{\color{hblue}34}\hspace{-1pt}){\color{hred}71}\!\big)\tncap(\hspace{-1pt}56\hspace{-1pt})7\!\Big)\!\Big],\\[6pt]
N_1^{4}\!\!\equiv\!(\hspace{-1pt}13\hspace{-1pt})\,,\; 
N_2^{4}\!\!\equiv\!\!\Big[\!\Big(\!6\hspace{-1pt}\big(\hspace{-1pt}{\color{hred}56}(\hspace{-1pt}{\color{hblue}34}7\hspace{-1pt})\tncap\big(\hspace{-1pt}(\hspace{-1pt}234\hspace{-1pt})\tncap(\hspace{-1pt}71\hspace{-1pt})2\hspace{-1pt}\big)\!\big)\hspace{0pt}\tncap(\hspace{-1pt}71\hspace{-1pt})\!\Big)\mi\hspace{1pt}\!\Big(\!6\hspace{-1pt}\big(\hspace{-1pt}{\color{hblue}34}(\hspace{-1pt}{\color{hred}56}7\hspace{-1pt})\tncap\big(\hspace{-1pt}(\hspace{-1pt}234\hspace{-0pt})\tncap(\hspace{-1pt}71\hspace{-1pt})2\hspace{-1pt}\big)\!\big)\hspace{-1pt}\tncap(\hspace{-1pt}71\hspace{-1pt})\!\Big)\!\Big].
\end{array}\hspace{-50pt}\label{heptagon_b_ladder_tensor_numerators}}

\paragraph{Non-MHV Octagon Ladder Numerator Details}~\\[-14pt]

The octagon ladder integrals,
\eq{\fig{-29.67pt}{1}{octagon_a_ladder_integral},\label{octagon_a_ladder_appendix}}
are identical to the heptagons in dual-momentum space:
\eq{\hspace{-20pt}I_{8,\text{A}}^{i,(L)}\equiv\!\!\int\!\!d^{4L}\vec{\ell}\frac{\x{\ell_1}{N_1^i}\x{1}{4}^{L-2}\x{\ell_L}{N_2^i}}{\x{\ell_1}{2}\x{\ell_1}{3}\big(\prod_{i=1}^L\x{\ell_i}{1}\x{\ell_i}{4}\big)\big(\x{\ell_1}{\ell_2}\cdots\x{\ell_{L-1}}{\ell_L}\big)\x{\ell_L}{6}\x{\ell_L}{7}}\,.\hspace{-20pt}\label{octagon_a_ladder_loop_space_definition}}
In this case, the numerators are given by
\vspace{-24pt}\eq{\hspace{-67pt}\begin{array}{@{}l@{}}
~\\[12pt]N_1^{1}\!\!\equiv\!\!(\hspace{-1pt}812\hspace{-1pt})\tncap(\hspace{-1pt}234\hspace{-1pt})\,,\; 
N_2^{1}\!\!\equiv\!\!\Big[\!\Big(\!\big({\color{hred}34}(6{\color{hblue}81})\tncap(13)\big)\tncap(567)\!\Big)\mi\hspace{1pt}\!\Big(\!\big({\color{hblue}81}(6{\color{hred}34})\tncap(13)\big)\tncap(567)\!\Big)\!\Big],\\[6pt]
N_1^{2}\!\!\equiv\!\!(\hspace{-1pt}812\hspace{-1pt})\tncap(\hspace{-1pt}234\hspace{-1pt})\,,\; 
N_2^{2}\!\!\equiv\!\!\Big[\!\Big(\!6\big(13({\color{hred}81})\tncap(567)\big)\tncap({\color{hblue}34})\!\Big)\mi\hspace{1pt}\!\Big(\!6\big(13({\color{hblue}34})\tncap(567)\big)\tncap({\color{hred}81})\!\Big)\!\Big],\\[6pt]
N_1^{3}\!\!\equiv\!(\hspace{-1pt}13\hspace{-1pt})\,,\; 
N_2^{3}\!\!\equiv\!\!\Big[\!\Big(\hspace{-1pt}\!\big({\color{hred}34}(\hspace{-1pt}6{\color{hblue}81}\hspace{-1pt})\tncap\big(\hspace{-1pt}2(\hspace{-1pt}81\hspace{-1pt})\tncap(\hspace{-1pt}234\hspace{-1pt})\!\big)\!\big)\tncap(\hspace{-1pt}567\hspace{-1pt})\!\hspace{-1pt}\Big)\mi\hspace{1pt}\!\Big(\hspace{-1pt}\!\big({\color{hblue}81}(\hspace{-1pt}6{\color{hred}34}\hspace{-1pt})\tncap\big(\hspace{-1pt}2(\hspace{-1pt}81\hspace{-1pt})\tncap(\hspace{-1pt}234\hspace{-1pt})\!\big)\!\big)\tncap(\hspace{-1pt}567\hspace{-1pt})\!\Big)\!\Big],\\[6pt]
N_1^{4}\!\!\equiv\!(\hspace{-1pt}13\hspace{-1pt})\,,\; 
N_2^{4}\!\!\equiv\!\!\Big[\!\Big(\!6\hspace{-0pt}(\hspace{-1pt}{\color{hred}34}\hspace{-1pt})\tncap\big(\hspace{-1pt}2\hspace{-1pt}(\hspace{-1pt}{\color{hblue}81}\hspace{-1pt})\tncap(\hspace{-1pt}567\hspace{-1pt})\,(\hspace{-1pt}81\hspace{-1pt})\tncap(\hspace{-1pt}234\hspace{-1pt})\!\big)\!\hspace{-1pt}\Big)\mi\hspace{1pt}\!\hspace{0.5pt}\Big(\!6\hspace{-0pt}(\hspace{-1pt}{\color{hblue}81}\hspace{-1pt})\tncap\big(\hspace{-1pt}2\hspace{-1pt}(\hspace{-1pt}{\color{hred}34}\hspace{-1pt})\tncap(\hspace{-1pt}567\hspace{-1pt})\,(\hspace{-1pt}81\hspace{-1pt})\tncap(\hspace{-1pt}234\hspace{-1pt})\!\big)\!\hspace{-1pt}\Big)\!\Big].
\end{array}\hspace{-50pt}\label{octagon_a_ladder_tensor_numerators}}

\newpage\vspace{-6pt}
\section{Comparing Various (Cluster) Coordinates}\label{appendix:cluster_chart_comparisions}%
\vspace{-6pt}

In \mbox{section \ref{subsec:momentum_twistors}}, we made reference to the canonical (i.e., lex-min bridge-constructed) chart for the top-dimensional configuration of momentum twistors, understood as a subspace of $G_+(4,n)$: 
\eq{Z^{(n)}_{\text{seed}}\bigger{\Leftrightarrow}\fig{-42.185pt}{1}{top_cell_seed_edge_chart}.\label{plabic_graph_for_seed_chart}}
This graph (with different labels attached to the same edge variables) is exactly the one used to construct the output of {\tt permToMatrix[Range[n]+4]} in the {\sc Mathematica} package {\tt positroids} \cite{Bourjaily:2012gy}. Upon relabeling and setting the edge variables associated with the external faces of the graph to one, this matrix is identical to 
\eq{Z^{(n)}_{\text{seed}}(\vec{e})\equiv\!\raisebox{-0pt}{$\left(\raisebox{54pt}{}\right.$}\begin{array}{@{$\;$}c@{$\;$}c@{$\;\;$}c@{$\;\;$}c@{$\;\;\;$}c@{$\;$}c@{$\;$}c@{$\;$}}z_1&z_2&z_3&z_4&\cdots&z_{a\in[4,n]}&\cdots\\[-1pt]\hline~\\[-14pt]
1&\displaystyle\hspace{-2pt}\sum_{4<k\leq n}\hspace{-6pt}e_k^3&\displaystyle\hspace{-2pt}\sum_{4<j<k\leq n}\hspace{-10pt}e_j^2e_k^3&\displaystyle\hspace{-2pt}\sum_{4<i<j<k\leq n}\hspace{-16pt}e_i^1e_j^2e_k^3&\cdots&\displaystyle\hspace{-2pt}\sum_{a<i<j<k\leq n}\hspace{-16pt}e_i^1e_j^2e_k^3&\cdots\\
0&1&\displaystyle\hspace{-2pt}\sum_{4<j\leq n}\hspace{-6pt}e_j^2&\displaystyle\hspace{-2pt}\sum_{4<i<j\leq n}\hspace{-10pt}e_i^1e_j^2&\cdots&\displaystyle\hspace{-2pt}\sum_{a<i<j\leq n}\hspace{-10pt}e_i^1e_j^2&\cdots\\
0&0&1&\displaystyle\hspace{-2pt}\sum_{4<i\leq n}\hspace{-6pt}e_i^1&\cdots&\displaystyle\hspace{-2pt}\sum_{a<i\leq n}\hspace{-6pt}e_i^1&\cdots\\
0&0&0&1&\cdots&1&\cdots\\~\end{array}\raisebox{-0pt}{$\left.\raisebox{54pt}{}\right)$}\,.\label{seed_edge_parameterization_for_top_cell}}
In the matrix given above, we have made reference to $e_5^i\!\equiv\!1$ for notational compactness. If the edge variables $e_a^i$ are viewed as {\it coordinates}, they correspond to the chart\\[-6pt]
\eq{\begin{array}{@{}c@{}}\displaystyle e_a^1\equiv\frac{\ab{1234}\ab{1235}\ab{12a\mi\!1a}}{\ab{123a\mi\!1}\ab{123a}\ab{1245}}\,,\hspace{50pt}e_a^2\equiv\frac{\ab{123a\mi\!1}\ab{1245}\ab{1a\mi\!2a\mi\!1a}}{\ab{12a\mi\!2a\mi\!1}\ab{12a\mi\!1a}\ab{1345}}\,,\\[10pt]
\displaystyle e_a^3\equiv\frac{\ab{12a\mi\!2 a\mi\!1}\ab{1345}\ab{a\mi\!3a\mi\!2a\mi\!1a}}{\ab{1a\mi\!3a\mi\!2a\mi\!1}\ab{1a\mi\!2a\mi\!1a}\ab{2345}}\,.\end{array}\label{seed_edge_coordinates_for_top_cell}}
This is easy to confirm by computing determinants of the matrix (\ref{seed_edge_parameterization_for_top_cell}) directly. 

The {\it face variables} of a plabic graph are examples of so-called cluster $\mathcal{X}$-coordinates \cite{FG2,FG3,ArkaniHamed:2012nw}, and are associated with the nodes of the dual graph. For instance, the quiver
\eq{\raisebox{36.5pt}{\xymatrix @R=14pt @C=20pt{f_6^1\ar[dr]&\ar[l]f_7^1\ar[dr]&\ar[l]f_8^1\ar[dr]&\ar[l]\phantom{f_9^1}\hspace{-12pt}\cdots\ar[dr]&\ar[l]f_n^1\\
f_6^2\ar[dr]\ar[u]&\ar[l]f_7^2\ar[u]\ar[dr]&\ar[l]f_8^2\ar[dr]\ar[u]&\ar[l]\phantom{f_9^2}\hspace{-12pt}\cdots\ar[dr]&\ar[l]f_n^2\ar[u]\\
f_6^3\ar[u]&f_7^3\ar[l]\ar[u]&\ar[l]\ar[u]f_8^3&\ar[l]\phantom{f_9^3}\hspace{-12pt}\cdots&\ar[l]\ar[u]f_n^3}}\,\label{seed_x_quiver}}
is dual to (\ref{plabic_graph_for_seed_chart}), and comes equipped with $3(n\mi5)$ such variables. These variables are related to boundary measurements and are defined in terms of the edge variables (\ref{seed_edge_coordinates_for_top_cell}) in the obvious way \mbox{\cite{ArkaniHamed:2012nw}}:
\eq{f_a^i\equiv e^i_a/e^i_{a-1}\,.\label{face_via_edges_rule}}

A possible (but serious) source of confusion is that other $\mathcal{X}$-type cluster coordinates can be associated with (the faces of) the {\it same} plabic graph (\ref{plabic_graph_for_seed_chart}) and the {\it same} quiver diagram (\ref{seed_x_quiver}). In particular, in the context of symbology (e.g.\ \mbox{refs.\ \cite{Golden:2013xva,Golden:2014xqa,Drummond:2014ffa,Dixon:2016nkn,DelDuca:2016lad,Drummond:2017ssj}}), the more commonly used $\mathcal{X}$-coordinates can be thought of as derived from an $\mathcal{A}$-type quiver diagram. This $\mathcal{A}$-coordinate quiver can be constructed from~\eqref{plabic_graph_for_seed_chart} using left-right paths, as described in ref.~\cite{ArkaniHamed:2012nw}:
\eq{\raisebox{44pt}{$\xymatrix @M=1pt @R=14pt @C=20pt{{\color{hred}\sb{5678}}\ar[dr]&\ar@{.>}[l]{\color{hred}\sb{6789}}\ar[dr]&\ar@{.>}[l]\fwbox{18pt}{\cdots}\ar[dr]&\ar@{.>}[l]{\color{hred}\sb{n\mi\!1 n 12}}\ar[dr]&\ar@{.>}[l]{\color{hred}\sb{n 123}}\ar@{.>} [ddddr] <8pt>  \\
{\color{hred}\sb{4567}}\ar[dr]\ar@{.>}[u]&\ar[u]\ar[l]\sb{4\,678}\ar[dr]&\ar[l]\phantom{f_9^2}\hspace{-12pt}\cdots\ar[dr]&\ar[l]\sb{4\,n\mi\!1n1}\ar[u]\ar[dr]&\ar[l]\sb{4\,n12}\ar[u]&~\\
{\color{hred}\sb{3456}}\ar@{.>}[u]\ar[dr]&\ar[l]\sb{34\,67}\ar[dr]\ar[u]&\ar[l]\phantom{f_9^2}\hspace{-12pt}\cdots\ar[dr]&\ar[l]\sb{34\,n\mi\!1n}\ar[u]\ar[dr]&\ar[l]\ar[u]\sb{34\,n1}&~\\
{\color{hred}\sb{2345}}\ar@{.>}[u]\ar@{.>}[rrrrrd]<-4pt>&\ar[l]\sb{234\,6}\ar[u]&\ar[l]\phantom{f_9^2}\hspace{-12pt}\cdots&\ar[l]\ar[u]\sb{234\,n\mi\!1}&\ar[u]\ar[l]\sb{234\,n}&~\\&&&&&\ar@<-0pt>[ul] \raisebox{0pt}{\hspace{5pt}${\color{hred}\sb{1234}}$}\,.
}$}\label{a_quiver_for_seed}}
(Note that this isn't the exact $\mathcal{A}$-coordinate seed cluster usually seen in the symbology literature, but is related to that more familiar seed by simply shifting the $\mathcal{A}$-labels.) From this quiver, one can construct ${\cal X}$-coordinates at each non-frozen node by assigning the product of all $\mathcal{A}$-coordinates associated with outgoing arrows to the numerator, and the product of all $\mathcal{A}$-coordinates associated with incoming arrows to the denominator. (The frozen nodes of (\ref{a_quiver_for_seed}) are indicated in red.) 

Although far from obvious, it was shown in \mbox{ref.\ \cite{positroidTwists}} that these two representations are the same; they merely correspond to coordinates on different spaces. Specifically, there exists a (in this case \mbox{left-)twist} map \mbox{$\tau_L\!:\!Z\!\mapsto\!\widetilde{Z}$} letting us define
\eq{\widetilde{Z}\equiv\tau_L\big(Z\big)\quad\text{with}\quad\sb{abcd}\equiv\det\{\tilde{z}_a,\tilde{z}_b,\tilde{z}_c,\tilde{z}_d\}\,.\label{twist_operation}}
Then one can easily check that
\eq{\begin{split}f_a^1\,&\fwboxL{330pt}{=\frac{\ab{123a\mi\!2}\ab{12a\mi\!1a}}{\ab{123a}\ab{12a\mi\!2a\mi1}}}\fwboxR{0pt}{=\hspace{15pt}\frac{\sb{34a\pl\!1a\pl\!2}\sb{4a\mi\!1aa\pl\!1}\sb{aa\pl\!1a\pl\!2a\pl\!3}}{\sb{34aa\pl\!1}\sb{4a\pl\!1a\pl\!2a\pl\!3}\sb{a\mi\!1aa\pl\!1a\pl\!2}}}\,,\\
f_a^2\,&\fwboxL{330pt}{=\frac{\ab{123a\mi\!1}\ab{12a\mi\!3a\mi\!2}\ab{1a\mi\!2a\mi\!1a}}{\ab{123a\mi\!2}\ab{12a\mi\!1a}\ab{1a\mi\!3a\mi\!2a\mi\!1}}}\fwboxR{0pt}{=\hspace{-1pt}\frac{\sb{234a\pl\!1}\sb{34a\mi\!1a}\sb{4aa\pl\!1a\pl\!2}}{\sb{234a}\sb{34a\pl\!1a\pl\!2}\sb{4a\mi\!1aa\pl\!1}}}\,,\\
f_a^3\,&\fwboxL{330pt}{=\frac{\ab{12a\mi\!2a\mi\!1}\ab{1a\mi\!4a\mi\!3a\mi\!2}\ab{a\mi\!3a\mi\!2a\mi\!1a}}{\ab{12a\mi\!3a\mi\!2}\ab{1a\mi\!2a\mi\!1a}\ab{a\mi\!4a\mi\!3a\mi\!2a\mi\!1}}}\fwboxR{0pt}{=\hspace{8pt}\frac{\sb{234a\mi\!1}\sb{34aa\pl\!1}}{\sb{234a\pl\!1}\sb{34a\mi\!1a}}}\,.
\end{split}\label{seed_face_coordinates_with_and_without_twist}}

It may be helpful to give one concrete illustration of how these charts are related by twists. Consider the case of six particles:
\eq{Z^{(6)}_{\text{seed}}\equiv\!\raisebox{-0pt}{$\left(\raisebox{28pt}{}\right.$}\begin{array}{@{}c@{$\;\;$}c@{$\;\;$}c@{$\;\;$}c@{$\;\;$}c@{$\;\;$}c@{}}1&1\pl f_6^3&f_6^3&0&0&0\\[-2pt]
0&1&1\pl f_6^2&f_6^2&0&0\\[-2pt]
0&0&1&1\pl f_6^1&f_6^1&0\\[-2pt]
0&0&0&1&1&1\end{array}\raisebox{-0pt}{$\left.\raisebox{28pt}{}\right)$}}
with
\eq{f_6^1\equiv\frac{\ab{1234}\ab{1256}}{\ab{1236}\ab{1245}}\,,\quad f_6^2\equiv\frac{\ab{1235}\ab{1456}}{\ab{1256}\ab{1345}}\,,\quad f_6^3\equiv\frac{\ab{1245}\ab{3456}}{\ab{1456}\ab{2345}}\,,}
where we have used the fact that the edge coordinates are identical to the faces for six particles. The twist map $\tau_L$ results in a matrix (see \mbox{ref.\ \cite{positroidTwists}})
\eq{\widetilde{Z}^{(6)}\equiv\tau_L\big(Z^{(6)}_{\text{seed}}\big)=\!\raisebox{-0pt}{$\left(\raisebox{28pt}{\!}\right.$}\begin{array}{@{}c@{$\;\;$}c@{$\;\;$}c@{$\;\;$}c@{$\;\;$}c@{$\;\;$}c@{}}1&1&0&0&f_6^1f_6^2&0\\[-2pt]0&1&1\pl f_6^1&f_6^1&0&0\\[-2pt]
1\pl f_6^3&f_6^3&0&0&0&\mi1\\[-2pt]
0&0&1&1&1&0\end{array}\raisebox{-0pt}{$\left.\raisebox{28pt}{}\right)$},}
from which it is easy to see that 
\eq{f_6^1=\frac{\sb{1236}\sb{1456}}{\sb{1256}\sb{1346}}\,,\quad f_6^2=\frac{\sb{1246}\sb{3456}}{\sb{1456}\sb{2346}}\,,\quad f_6^3=\frac{\sb{1346}\sb{2345}}{\sb{1234}\sb{3456}}}
as a special case of (\ref{seed_face_coordinates_with_and_without_twist}).

\newpage

\providecommand{\href}[2]{#2}\begingroup\raggedright\endgroup

\end{document}